\documentclass[manuscript]{aastex701}

\begin{document}

\title{Balmer Absorption in Iron Low-Ionization Broad Absorption Line
  Quasars}

\author[orcid=0000-0002-3809-0051,sname='Leighly']{Karen M. Leighly}
\affiliation{Homer L.\ Dodge Department of Physics and Astronomy, The
  University of Oklahoma, 440 W.\ Brooks St., Norman, OK 73019, USA}
\email[show]{leighly@ou.edu}
\author{Sarah C.\ Gallagher}
\affiliation{Department of Physics \& Astronomy, The University of
  Western Ontario, London, ON, N6A 3K7, Canada}
\affiliation{Institute for Earth and Space Exploration, The
  University of Western Ontario, London, ON, N6A 3K7, Canada}
\email{sgalla4@uwo.ca}
\author[0000-0002-3173-1098]{Hyunseop Choi}
\affiliation{D\'{e}partement de Physique, Universit\'{e} de
  Montr\'{e}al, Succ.\ Centre-Ville, Montr\'{e}al, Qu\'{e}bec, H3C 3J7,
  Canada}
\email{hyunseop.choi@montreal.ca}
\author{Donald M.\ Terndrup}
\affiliation{Homer L.\ Dodge Department of Physics and Astronomy,
  The
  University of Oklahoma, 440 W.\ Brooks St., Norman, OK 73019, USA}
\affiliation{Department of Astronomy, The Ohio State University, 140
  W. 18th Ave., Columbus, OH 43210, USA}
\email{terndrup.1@osu.edu}
\author{Julianna R. Voelker}
\affiliation{Homer L.\ Dodge Department of Physics and Astronomy,
  The
  University of Oklahoma, 440 W.\ Brooks St., Norman, OK 73019, USA}
\email{juliannarvoelker@gmail.com}
\author{Gordon T.\ Richards}
\affiliation{Department of Physics, Drexel University, 32 S. 32nd St.,
  Philadelphia, PA 19104, USA}
\email{gtr25@drexel.edu}
\author{Leah K.\ Morabito}
\affiliation{Centre for Extragalactic Astronomy, Department of
  Physics, Durham University, Durham DH1 3LE, UK}
\affiliation{Institute for Computational Cosmology, Department of
  Physics, Durham University, Durham DH1 3LE, UK} 
\email{leah.k.morabito@durham.ac.uk}

\begin{abstract}

While \ion{C}{4} is the most common absorption line in Broad
Absorption Line Quasar spectra, Balmer absorption lines are among the
rarest.  We present analysis of Balmer absorption in a sample of
fourteen iron low-ionization BAL quasars (FeLoBALQs); eight are new
identifications.  We measured velocity offset, width, and apparent
optical depth.   The partial covering ubiquitous in BAL quasar spectra
alters the measured Balmer optical depth ratios; taking that into
account, we estimated the true H($n=2$) column density.
We found the anticipated correlation between Eddington ratio and
outflow speed, but it is weak in this sample because nearly all of the
objects have the low outflow speeds characterizing loitering
outflow FeLoBAL quasars \citep{choi22}, objects that are also found to
have low accretion rates \citep{leighly22, choi22b}. Measures of
$dN/dv$, the differential column density with respect to the outflow
speed, are anticorrelated with the luminosity and Eddington ratio:
the strongest absorption is observed at the lowest speeds in the
lowest luminosity objects.  The absorption line width is correlated
with $\alpha_{oi}$, the $F_\lambda$ point-to-point slope between 5100\AA\/ and
3$\mu \rm m$.  This parameter is
strongly correlated with 
the Eddington ratio among low-redshift quasars
\citep{leighly24}. Balmer absorption lines have been recently found in 
the spectra of Little Red Dots (LRDs), a class of high-redshift
objects discovered by JWST.  We note suggestive similarities between
LRDs and FeLoBAL quasars in the emission line shape, the
presence of steep reddening and a scattered blue continuum, the lack
of hot dust emission, and X-ray weakness.  
  
\end{abstract}



\section{Introduction} \label{sec: intro}

Broad absorption lines are found in about 10--26\% of optically selected
quasars  \citep{tolea02, hf03, reichard03, trump06, knigge08,
  gibson09}.  The absorption lines are sometimes very broad, with
velocity widths of up to $10,000 \rm\, km\, s^{-1}$ and strongly 
blueshifted \citep[e.g.,][]{baskin15} indicating outflow velocities
that sometimes exceed 0.1$c$.  Broad 
absorption line quasars (BALQs) are classified according to the lines
that appear in their rest-UV spectra.  High-ionization BALQs show
absorption from \ion{C}{4}, \ion{N}{5}, \ion{O}{6}, and Ly$\alpha$;
depending on the column density and ionization parameter of the
outflow, they can also have \ion{Si}{4}, \ion{S}{4}, and \ion{P}{5} in
their spectra.  Low-ionziation BALQs (LoBALQs) have all of these
lines, as well as \ion{Mg}{2} and \ion{Al}{3}.  The presence of these
lines indicates that the column density of the outflow nearly reaches
the hydrogen ionization front.  Finally, iron low-ionization BALQs
(FeLoBALQs) have all of those lines as well as absorption from
\ion{Fe}{2}.

Recently, we  used the novel spectral synthesis code {\tt SimBAL}
\citep{leighly22} to analyze a sample of 50 low-redshift FeLoBAL
quasars \citep{choi22}, increasing the number of well-studied
FeLoBALQs in the literature by a factor of five.  Although they are
rare, comprising only about 0.3\% of optically selected quasars
\citep{trump06}, FeLoBAL quasars have the highest column densities
among BAL quasars and can drive some of the most massive and energetic
quasar winds observed  \citep[e.g.,][]{choi20}. These quasars also
typically show reddened spectral energy distributions, which, together
with their powerful winds, suggest that they may represent a
transitionary “blow-out" evolutionary stage, during which a quasar
actively expels gas and dust, transforming from a heavily shrouded
ultraluminous, infrared galaxy-type (ULIRG) object to a blue quasar
\citep[e.g.,][]{farrah12}.  We found that the outflows lie at a large
range of distances   from the central engine, from ~1 pc (near the
torus), to kiloparsecs (in the host galaxy) \citep{choi22}. We found
that the accretion rate distribution of FeLoBAL quasars is roughly
bimodal, with a low accretion rate branch centered at
$L_\mathrm{Bol}/L_\mathrm{Edd} \sim -1.0$ and a high accretion rate
branch with $L_\mathrm{Bol}/L_\mathrm{Edd} \sim 0.0$; in contrast, the
accretion rate distribution of an unabsorbed comparison sample is
strongly peaked at $L_\mathrm{Bol}/L_\mathrm{Edd} \sim -0.5$
\citep{leighly22}.  We found that the outflow properties are linked to
the accretion rate.  For high accretion rate objects, the outflow
speed\footnote{Throughout this paper, we use the convention that
outflow velocities are negative in sign, and the magnitude of the
velocity is referenced as the speed.} decreases with radius as
expected in radiative acceleration scenarios because the radiation
intensity is largest near the central engine.   However,
\citet{choi22} found significant population of 
outliers from this relationship.  Called loitering outflow FeLoBAL
quasars, they are characterized by low outflow velocity
($|V_\mathrm{off}| < 2000 \rm \, km\, s^{-1}$), and location near the
central engine ($\log R < 1$ [pc]); we note that the 
class includes objects where the gas is inflowing.  Loitering
outflow FeLoBAL quasars are also characterized by low accretion rates 
\citep{leighly22,   choi22b}.  Finally, the photometry and near-IR
variability properties were analyzed; notably, we found evidence that
the spectral energy distributions of  the loitering outflow FeLoBAL
quasars show little evidence for reddening and a lack of hot dust
emission \citep{leighly24}.  The results from these four papers
yielded ample evidence that low-redshift FeLoBALQs are distinct from
unabsorbed objects, prompting development of a scenario for their
character and evolution whereby the high-accretion-rate
FeLoBAL quasars are radiating powerfully enough to drive a thick,
high-velocity outflow, quasars with
intermediate accretion rates may have an outflow, but it is not
sufficiently thick to include \ion{Fe}{2} absorption, and low
accretion-rate FeLoBAL outflows originate in absorption in a failing
torus, no longer optically thick enough to reprocess radiation into
the near-IR  \citep{leighly24}. 

Recently, JWST has revealed a specific class of low-luminosity
high-redshift object that are different from high-redshift quasars.  
Little Red Dots (LRDs) are found to have peculiar V-shaped
spectra, red in the 
rest-frame optical band, and blue in the near-UV band.  Many of these
objects are thought to be Active Galactic Nuclei (AGN), as they have
broad Balmer emission 
lines. A number of LRDs show narrow and mildly blueshifted Balmer
absorption lines \citep[e.g.,][]{deugenio25, kocevski25, matthee24};
the fraction of objects with these lines may be as large as 20\% 
\citep[e.g.,][]{inayoshi25}.

BAL quasars are another class of extragalactic object that has
occasionally been found to have Balmer line absorption. Most of these
objects are FeLoBAL quasars \citep{hall02, 
  aoki06, aoki10, zhang15, schulze18}, although there are some reports
of weak absorption in spectra from non-BAL quasars when the
signal-to-noise ratio is excellent \citep{hutchings02, wang08, ji12,
  wang15}.  In a photoionized gas, Balmer absorption requires fairly
extreme conditions including a high density \citep[e.g.,][]{leighly11},
high column density, and high ionization parameter, all of which work 
together to keep the hydrogen $n=2$ population elevated enough to yield
observable absorption.  Given that FeLoBAL quasars show the highest
absorption column densities among BAL quasars, it is not surprising
that they  show the largest equivalent Balmer absorption lines.  In
fact, \citet{choi22} used the {\tt SimBAL} results to predict which of
their sample of low-redshift FeLoBALs should have Balmer absorption
lines (Fig.\ 16 in that paper). 

To follow up our results of analysis of low-redshift FeLoBAL quasars,
we have initiated a project to perform {\tt SimBAL} analysis on a
sample of high-redshift objects (Choi et al.\ in prep.).  At the same
time, we are performing near-infrared spectroscopic observations using
Gemini GNIRS and APO Triplespec to see if the differences found in the
rest-frame optical spectra of low-redshift FeLoBALQs \citep{leighly22}
carry over to their more luminous high-redshift brethren.  In this
paper, we present nine of our new observations with Balmer
absorption in their spectra.  We supplement the sample with two spectra
from the literature \citep{schulze18}, and one from the Gemini archive
for a set of 12 objects that have H$\alpha$ absorption detected in
their band pass.  We also include two from
\citet{choi22, leighly22} that have H$\beta$ absorption but lack NIR 
spectra required to view H$\alpha$.  The final sample comprisies 14
objects.

\S\ref{sample} describes the sample and extraction of the
spectra. \S\ref{analysis} discusses the spectral fitting and
extraction of parameters.  \S\ref{correlations} includes the
correlation analysis.  \S\ref{discussion} discusses what we can learn
about the physical properties of FeLoBAL quasar absorption from the
results, and it also compares FeLoBAL quasar properties with those of
LRDs.  It turns out that there are similarities in the emission lines
and in the spectral energy distribution, including the lack of hot
dust emission, the extreme reddening and blue emission, and lack of
X-ray emission.  \S\ref{summary} summarizes the results.  Throughout
we use typical cosmological parameters (H$_0=69.32 \rm \, km\,
Mpc^{-1}\, s^{-1}$, $\Omega_c=0.233$, $\Omega_\Lambda=0.721$).  

\section{The Sample} \label{sample}

FeLoBAL quasars can be studied in ground-based optical spectra for
redshifts less than about 3 due to the plethora of \ion{Fe}{2}
absorption lines produced in the ionized gas.  We are performing {\tt
  SimBAL} analysis on a sample of $\sim 50$ high redshift quasars
\citep[][Choi et al. in prep.]{voelker21}; this sample has bolometric 
luminosities about 1 dex higher than our previously-studied
low-redshift sample (median $\log L_\mathrm{Bol} \sim 47.1$ versus
46.2 [erg s$^{-1}$]) described 
in \citet{choi22, choi22b, leighly22, leighly24}.  
Depending on the redshift, the H$\beta$ / \ion{Fe}{2} / [\ion{O}{3}]
spectral region can fall in the near-infrared bands accessible from
the ground; specifically, for $1.3
< z < 1.6$, this region falls in the J band, and for $2.1 < z < 2.6$,
it falls in the H band.   Given what
we learned about FeLoBAL quasars from the low redshift sample, we 
targeted objects that had a wide range of \ion{Fe}{2} morphologies,
including objects with a moderate amount of excited state \ion{Fe}{2}
\citep[e.g.,][]{lucy14}, overlapping trough quasars
\citep[e.g.,][]{hall02}, and the loitering outflow FeLoBALQs,
characterized by very high excitation \ion{Fe}{2} lines but low
outflow velocities first identified by \citet{choi22}.  We performed
observations using TripleSpec \citep{wilson04} on the Astrophysical
Research Consortium (ARC) 3.5 meter telescope, and using GNIRS
\citep{elias06, elias09} on the Gemini North telescope.  Two spectra
were  published by \citet{schulze18}.  A few spectra were
obtained from the Gemini archive.

From the full sample of more than thirty objects, twelve included
Balmer absorption in their spectra (Table~\ref{table1}).  We also
include two objects previously presented in \citet{choi22} and
\citet{leighly22}; we present analysis of their Balmer absorption
lines here.  The analysis of the full sample,  and well as the
 {\tt SimBAL} analysis, focusing principally on the \ion{Fe}{2}
 absorption in the UV band, will be presented elsewhere (Leighly 
et al.\ in prep., Choi et al.\ in prep.)

The Gemini GNRIS data were reduced using {\tt PypeIt}
\citep{prochaska20}.  The APO TripleSpec data were reduced using
{\tt TripleSpecTool}, a modification of {\tt SpexTool} \citep{cushing04, vacca03}.  

Of our fourteen objects, six were already known to have Balmer
absorption (Table~\ref{table1}), and eight are new detections, having
never been observed in the NIR before.
Several other FeLoBALQs are  
known to have Balmer absorption, including SDSS~J125942.80$+$121312.6
\citep[$z=0.748$,][]{hall07} and SDSS~J152350.42$+$391405.2
\citep[$z=0.661$,][]{zhang15}.  SDSS~J122826.79$+$100532.2 has Balmer
absorption, and may have low-covering-fraction FeLoBAL absorption
\citep{li21}.  Objects that are not FeLoBAL quasars have 
been found to have Balmer absorption, but generally it is very weak
and would be difficult to detect without the presented excellent
signal-to-noise ratio spectra.  These include
SDSS~J102839.11$+$450009.4 \citep[$z=0.583$,][]{wang08},
LBQS~1206$+$1052 \citep[$z=0.395$,][]{ji12}, 
NGC~4151\citep[$z=0.0034$,][]{hutchings02}, and  SDSS~J112611.63+425246
\citep[z=0.156,][]{wang15}.  A difference between our
  investigation and these examples is that while their discoveries
  were serendipious and limited to analysis of one or two
  objects, our target choices were predicated by our experience with
  FeLoBAL spectra, and our analysis focuses on the properties of the
  population.

Continuum properties were measured from archival photometry of the
objects.  For the observed frame optical band, we compiled SDSS
\citep{blanton17} and Pan-STARRS \citep{chambers16} photometry,
prioritizing Pan-STARRS when available.  For the observed frame
near-infrared band, we compiled 2MASS \citep{skrutskie06} and  UKIDSS
\citep{lawrence07} photometry, prioritizing UKIDSS photometry 
when available.   Seven objects did not have archival near-IR
photometry.   All objects had WISE photometry \citep{wright10}. 

The spectra and photometry were corrected for Galactic reddening using
the \citet{ccm88} extinction curves.  We had only SDSS spectra for two
of the objects (SDSS~1125$+$0029 and SDSS~1644$+$5307); their
redshifts were estimated using [\ion{O}{2}] \citep{choi22}.  Redshift
estimation for higher luminosity objects is a challenge since even
low-ionization lines that make the best redshift indicators show
evidence for outflows \citep[e.g.][]{zg14}.  For the remaining 12
objects, we uniformly used the peak value of H$\alpha$ for
consistency.

\movetabledown=0.5in
\begin{longrotatetable}
\begin{deluxetable*}{lCCccccc}
\tabletypesize{\scriptsize}
\tablecaption{Sample \label{table1}}
\tablehead{
  \colhead{SDSS Object Name} & \colhead{Redshift}\tablenotemark{a} & \colhead{H
    magnitude\tablenotemark{b}} & \colhead{Telescope \&  Instrument} & \colhead{Date
    Observed\tablenotemark{c}} & \colhead{Exposure Time} & \colhead{Data Identifier} &
  \colhead{Reference\tablenotemark{d}}  \\
& & & & & (seconds) & \\}
\startdata
080202.69+140315.1 & 2.143 & 16.05 & Gemini  GNIRS & 2023-Nov-04 &
3096.0 & GN-2023B-Q-239-39 & \nodata \\
083942.11+380526.4 & 2.318 & 16.28  & Gemini  GNIRS & 2023-Oct-29 &
3096.0 & GN-2023B-Q-239-48  & Aoki 2006 \\
085910.40+423911.3 & 1.497  & 15.38 & Palomar 200 inch / Triplespec
& \nodata  &  \nodata & \nodata &  Schulze et al.\ 2018 \\
101927.37+022521.3 & 1.364 & 15.22 & Palomar 200 inch / Triplespec
&  \nodata  &  \nodata &  \nodata & Schulze et al.\ 2018 \\
112526.12+002901.3 & 0.8636 & 16.22 & SDSS &  \nodata &  \nodata &  \nodata & Hall et al.\ 2002 \\
122933.32+262131.2 & 2.581 & 16.09 & ARC 3.5 TripleSpec & 2021-Feb-24 &
7680.0 & \nodata & \nodata \\
124452.49+583427.6 & 2.282 &  16.9\tablenotemark{e} & Gemini GNIRS & 2024-Feb-21 & 6192.0 &
GN-2024A-Q-142-23 & \nodata \\
143916.28+162858.5 & 2.218 & 16.6\tablenotemark{e}  & Gemini GNIRS & 2023-Jul-16 &  2880.0 &
GN-2023A-Q-230-31 & \nodata \\
160915.16+561943.2 & 1.33 &  16.6\tablenotemark{e} & Gemini GNIRS &
2023-Jun-30 & 3840.0 & 
GN-2023A-Q-230-62 & \nodata \\
162119.22+081950.7 & 2.572 &  17.0\tablenotemark{e}  & Gemini GNIRS & 2022-Sep-10 & 4800.0 &
GN-2022A-Q-327-46 & \nodata \\
162527.73+093332.8 & 2.290 & 16.35 & Gemini GNIRS & 2024-Feb-22 &
2048.0 & GN-2024A-Q-142-32 & \nodata \\
163515.87+143925.9 & 1.317 & 16.8\tablenotemark{e}  & Gemini GNIRS &
2022-Sep-03 & 3840.0 & 
GN-2022A-Q-237-153 & \nodata \\
164419.75+530750.4 & 0.7813 & 16.18 & SDSS &  \nodata  & \nodata  &
\nodata & Leighly et al.\ 2022 \\ 
172341.08+555340.5 & 2.108 & 15.30 & Gemini GNIRS & 2023-Jun-24 &
960.0 & GN-2023A-Q-103-53 & Aoki 2010 \\
\enddata
\tablenotetext{a}{The redshifts were estimated by us from the peak of
  the H$\alpha$ line when present; the source of the redshifts for the
two SDSS spectra is given in \citet{choi22}.  } 
\tablenotetext{b}{The infrared magnitudes listed are from 2MASS
  \citep{skrutskie06}  or UKIDSS \citep{lawrence07}, except when noted.} 
\tablenotetext{c}{The information is given only for spectra that we
  reduced; the remainder are publicly available in already-reduced
  form.  } 
\tablenotetext{d}{The reference that reported the discovery of Balmer absorption. } 
\tablenotetext{e}{The H magnitude values were estimated by
  interpolation between  optical (SDSS or PanSTARRS) and WISE values.}
\end{deluxetable*}
\end{longrotatetable}

\section{Data Analysis} \label{analysis}

\begin{figure*}[!t]
\epsscale{1.0}
\begin{center}
\includegraphics[width=6.5truein]{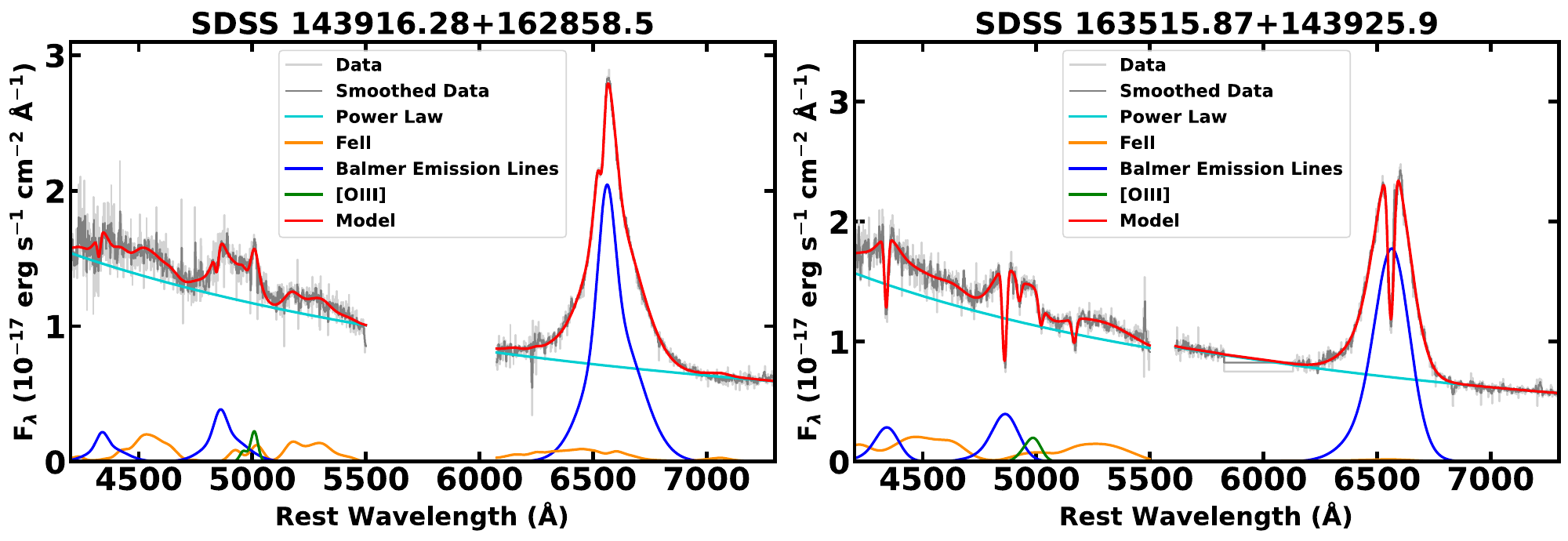}
\caption{Examples of the spectra and model fits for two of the
  fourteen spectra; the remainder are found in the online journal. We
  note that although the flux density units are given, the spectra are
  not reliably flux calibrated.
  {\it Left:} SDSS~1439$+$1628, the object with the lowest optical depth
  H$\alpha$ Balmer absorption line. {\it Right:} SDSS~1635$+$1439, the
  object with the largest optical depth H$\alpha$ Balmer absorption
  line.   The  complete figure set (14 images) is available in the
  online journal. 
   \label{specfit_plots}}
\end{center}
\end{figure*}

The spectra were analyzed using {\tt Sherpa} \citep{sherpa24}.  The
spectral analysis was hampered by the wide range of 
signal-to-noise ratios (SNR) of the spectra
(Fig.~\ref{specfit_plots}), due to both  the low flux of the objects,
and the atmospheric absorption features in the near-infrared spectra.
As outlined below, a simplified model and a  uniform approach was used
in order to obtain the same information from each spectrum regardless
of the spectral SNR.  An example of two of the spectra and model fits
are shown in Fig.~\ref{specfit_plots}, and the plots of the remainder
of the spectra are given in the online journal.  

\figsetstart
\figsetnum{1}
\figsettitle{Spectra and Model fits}
\figsetgrpstart

\figsetgrpnum{figurenumber.1}
\figsetgrptitle{080202.69+140315.1}
\figsetplot{sdss0802_model_fit.pdf}
\figsetgrpnote{Near infrared spectra and best fitting unconstrained
  model for SDSS~J0802+1403. }
\figsetgrpend

\figsetgrpnum{figurenumber.2}
\figsetgrptitle{083942.11+380526.4}
\figsetplot{sdss0839_model_fit.pdf}
\figsetgrpnote{Near infrared spectra and best fitting unconstrained
  model for SDSS~J0839+3805. }
\figsetgrpend

\figsetgrpnum{figurenumber.3}
\figsetgrptitle{085910.40+423911.3}
\figsetplot{sdss0859_model_fit.pdf}
\figsetgrpnote{Near infrared spectra and best fitting unconstrained
  model for SDSS~J0859+4239. }
\figsetgrpend

\figsetgrpnum{figurenumber.4}
\figsetgrptitle{101927.37+022521.3}
\figsetplot{sdss1019_model_fit.pdf}
\figsetgrpnote{Near infrared spectra and best fitting unconstrained
  model for SDSS~J1019+0225. }
\figsetgrpend

\figsetgrpnum{figurenumber.5}
\figsetgrptitle{112526.12+002901.3}
\figsetplot{sdss1125_model_fit.pdf}
\figsetgrpnote{Near infrared spectra and best fitting unconstrained
  model for SDSS~J1125+0029. }
\figsetgrpend

\figsetgrpnum{figurenumber.6}
\figsetgrptitle{122933.32+262131.2}
\figsetplot{sdss1229_model_fit.pdf}
\figsetgrpnote{Near infrared spectra and best fitting unconstrained
  model for SDSS~J1229+2621. }
\figsetgrpend

\figsetgrpnum{figurenumber.7}
\figsetgrptitle{124452.49+583427.6}
\figsetplot{sdss1244_model_fit.pdf}
\figsetgrpnote{Near infrared spectra and best fitting unconstrained
  model for SDSS~J1244+5834. }
\figsetgrpend

\figsetgrpnum{figurenumber.8}
\figsetgrptitle{143916.28+162858.5}
\figsetplot{sdss1439_model_fit.pdf}
\figsetgrpnote{Near infrared spectra and best fitting unconstrained
  model for SDSS~J1439+1628. }
\figsetgrpend

\figsetgrpnum{figurenumber.9}
\figsetgrptitle{160915.16+561943.2}
\figsetplot{sdss1609_model_fit.pdf}
\figsetgrpnote{Near infrared spectra and best fitting unconstrained
  model for SDSS~J1609+5619. }
\figsetgrpend

\figsetgrpnum{figurenumber.10}
\figsetgrptitle{162119.22+081950.7}
\figsetplot{sdss1621_model_fit.pdf}
\figsetgrpnote{Near infrared spectra and best fitting unconstrained
  model for SDSS~J1621+0819. }
\figsetgrpend

\figsetgrpnum{figurenumber.11}
\figsetgrptitle{162527.73+093332.8}
\figsetplot{sdss1625_model_fit.pdf}
\figsetgrpnote{Near infrared spectra and best fitting unconstrained
  model for SDSS~J1625+0933. }
\figsetgrpend

\figsetgrpnum{figurenumber.12}
\figsetgrptitle{163515.87+143925.9}
\figsetplot{sdss1635_model_fit.pdf}
\figsetgrpnote{Near infrared spectra and best fitting unconstrained
  model for SDSS~J1635+1439. }
\figsetgrpend

\figsetgrpnum{figurenumber.13}
\figsetgrptitle{164419.75+530750.4}
\figsetplot{sdss1644_model_fit.pdf}
\figsetgrpnote{Near infrared spectra and best fitting unconstrained
  model for SDSS~J1644+5307. }
\figsetgrpend

\figsetgrpnum{figurenumber.14}
\figsetgrptitle{172341.08+555340.5}
\figsetplot{sdss1723_model_fit.pdf}
\figsetgrpnote{Near infrared spectra and best fitting unconstrained
  model for SDSS~J1723+5553. }
\figsetgrpend

\figsetend

The continuum was modeled using a power law.  The H$\alpha$ and
H$\beta$ regions of the spectra were modeled with separate power laws.
The H$\alpha$ region was modeled typically between rest-frame 6000 and
7400\AA\/. A Lorenzian profile adequately modeled the Balmer emission
lines  in most 
cases, although two or three Gaussian profiles were used when the line
showed asymmetry.  The H$\alpha$ emission lines were strong and broad, and
 [\ion{N}{2}], if present, was too weak to be detected. Low-level
 \ion{Fe}{2} emission 
is observed shortward of H$\alpha$ \citep[e.g.,][]{vcjv04}; a model
constructed using the \ion{Fe}{2} line measurements from
\citet{vcjv04} was used, with the width of the lines set equal to the
Balmer Lorentzian width or the narrower of the two Gaussians
\citep[e.g.,][]{bg92}.

The H$\beta$ region was modeled between rest-frame 4200 and
5500\AA\/, and it was fit simultaneously with the H$\alpha$ region.
Both H$\beta$ and H$\gamma$ were modeled, although in two cases,
the H$\gamma$ region of the spectrum could not be fit due to poor SNR. 
The widths of these Balmer emission lines were constrained to be equal to the
H$\alpha$ line.  The \citet{kovacevic10} \ion{Fe}{2} model was used,
and again the width of the \ion{Fe}{2} lines was tied to the Balmer
line widths as described above.

The [\ion{O}{3}] emission feature was modeled with one or two
Gaussians depending 
on the spectrum; in some cases, [\ion{O}{3}] was not observed in the
spectrum, and the feature was not modeled.  As usual,
the two [\ion{O}{3}] components were constrained according to atomic
physics and their common kinematics.

The Balmer absorption lines were modeled using a Gaussian opacity
profile, and this model provided an excellent fit.  In most cases, one
component was sufficient; for three objects (SDSS~1125$+$0029,
SDSS~1244$+$5834, SDSS~1609$+$5619) two components were used.  We
acknowledge that the assumption of a Gaussian opacity profile may
obscure more complicated kinematics; however, we felt that a simple
model provided a fair treatment of all the spectra regardless of
signal-to-noise ratio.  Generally, we modeled H$\alpha$, H$\beta$, and
H$\gamma$ except in the cases where the spectrum did not include the
relevant bandpass or where the spectrum was exceptionally noisy.
While the positions and widths of the lines were constrained according
to atomic physics, our first model did not constrain the optical depth
ratios.  This model is referred to henceforth as the ``unconstrained''
model. 

\movetabledown=0.5in
\begin{longrotatetable}
\begin{deluxetable*}{l|CCCCC|cC|cC}
\tabletypesize{\scriptsize}
\tablecaption{Spectral Fitting Results \label{specfit_tab}}
\tablehead{
\colhead{Object Name} & \multicolumn{5}{c}{Unconstrained Model}
& \multicolumn{2}{c}{Constrained Model} & \multicolumn{2}{c}{Partial
  Covering Model} \\
& \colhead{FWHM} & \colhead{Velocity Offset\tablenotemark{a}} &
\colhead{H$\alpha$ $\tau_\mathrm{max}$} & \colhead{H$\beta$
  $\tau_\mathrm{max}$} & \colhead{H$\alpha$ Apparent Column
  } & \colhead{Statistically} &
\colhead{True Column } & \colhead{Best Partial} &
\colhead{Covering Fraction\tablenotemark{b}} \\
\colhead{} & \colhead{(km s$^{-1}$)} & \colhead{(km s$^{-1}$)} &
\colhead{} & \colhead{} & 
\colhead{Density ($10^{13}$ cm$^{-2}$)} & \colhead{Necessary Line} &
\colhead{Density (10$^{13}$
  cm$^{-2}$)} &  \colhead{Covering Model\tablenotemark{c}} \\ 
}
\startdata
080202.69+140315.1 & 1064 \pm 16 & -1360  \pm 5 & 0.367 \pm 0.005 &
0.128 \pm 0.006 &
3.72 \pm 0.08 & H$\alpha$ & 3.78 \pm 0.08 & \nodata & \nodata \\
083942.11+380526.4 & 539 \pm 7 & -237\pm 2.4 & 0.634 \pm 0.007 & 0.401
\pm 0.007 & 3.26
\pm 0.05 & H$\beta$ & 14.9 \pm 0.3 & FC+PL & \nodata \\
085910.40+423911.3 & 630^{+35}_{-33} & 780\pm 11 & 0.37 \pm 0.014 &
0.46 \pm 0.07 &  2.23
\pm 0.17 & H$\alpha$ & 2.11 \pm 0.14 & \nodata &  \nodata \\
101927.37+022521.3 &  1220^{+47}_{-51} & -1520 \pm 23 & 0.57 \pm 0.03
& 0.47^{+0.13}_{-0.12}  &
6.6 \pm 0.4 & H$\alpha$ & 6.6 \pm 0.4 & \nodata &  \nodata \\
112526.12+002901.3\tablenotemark{d}\tablenotemark{e} &
820^{+64}_{-600} & 430 \pm 45 &
\nodata & 
0.31\pm 0.03 & \nodata & H$\gamma$ & 11.3 \pm 2.5 & PC & 0.45^{+0.09}_{-0.06} \\
122933.32+583427.6 & 970^{+74}_{-69} & -1020\pm 30 & 0.33 \pm 0.02 &
0.021\pm 0.04 & 
3.0\pm 0.3  & H$\alpha$ & 3.1\pm 0.3 & \nodata & \nodata \\
124452.49+583427.6\tablenotemark{e} & 1330\pm 9 & -1720 \pm
4 & 0.435 \pm 0.006 & 0.287 
\pm 0.005 & 5.02 \pm 0.05 & 
H$\beta$ & 21.2 \pm 0.4 & FC+PL & \nodata \\
143916.28+162858.5 & 1127^{+30}_{-32} & -1170\pm 11 & 0.185 \pm 0.005
& 0.12 \pm 0.015 
& 1.99 \pm 0.08 & H$\alpha$ & 1.86\pm 0.07 & FC+PL & \nodata \\
160915.16+561943.2\tablenotemark{e} & 530 \pm 19 & -1460 \pm
14 & 0.56\pm 0.01 & 0.25
\pm 0.01 & 4.0
\pm 0.12 & H$\beta$ & 15.9 \pm 0.8 & FC+PL & \nodata \\
162119.22+081950.7 & 1410 \pm 41 & -4130 \pm 20 & 0.247 \pm 0.007 &
0.043 \pm 0.008 & 
3.33\pm 0.13 & H$\alpha$ & 3.33 \pm 0.13 & \nodata & \nodata \\
162527.73+093332.8 & 1010 \pm 110 & -6620 \pm 13 & 0.27 \pm 0.01 &
0.143 \pm 0.009 & 2.5
\pm 0.3 & H$\alpha$ & 2.5 \pm 0.34 & \nodata & \nodata \\
163515.87+143925.9 & 1235^{+11}_{-15} & -140 \pm 5 & 0.745 \pm 0.006 &
0.66 \pm 0.012 & 
8.8 \pm 0.1 & H$\gamma$ & 96 \pm 4 & FC+PL & \nodata  \\
164419.75+530750.4\tablenotemark{d} & 820 \pm 40 & -1560\pm 13 &
\nodata & 0.40 \pm 0.02 & \nodata &
H$\gamma$ & 14.0 \pm 1.2 & PC & 0.31^{+0.007}_{-0.025} \\
172341.08+555340.5\tablenotemark{f} & 1085^{+17}_{-16} & -5500\pm 6.4 & 0.47\pm
0.01 & 0.243 \pm 0.006 & 4.9 \pm
0.13 & H$\gamma$ & 12.1\pm 0.4 & P(L+C) & 0.21 \pm
0.004 \\ 
\enddata
\tablenotetext{a}{All of the objects meet the velocity criterion
  identifying a loitering outflow FeLoBAL quasar ($|V_\mathrm{off}| <
  2000\rm \, km\, s^{-1}$) except SDSS~J1621$+$0819, SDSS~J1625$+$0933,
and SDSS~J1723$+$5553.  Of these, only SDSS~J1723$+$5553 lacks the
strong very highly excited \ion{Fe}{2} absorption around 2500\AA\/
that signals a location near the torus \citep[e.g., Fig.\ 3][]{choi22}.}
\tablenotetext{b}{The step-function partial covering fraction in the
  case where the continuum is partially covered.  If the continuum is
  fully covered, the covering fraction is assumed to be 1.}
\tablenotetext{c}{No entry: the Balmer optical depth line ratios are
  statistically  consistent with the values required by atomic
  physics, and no partial covering is necessary.  P(L+C): partial
  covering of both the 
  emission lines and the continuum; FC+PL: full covering of the
  continuum, partial covering of the emission lines; PC: partial
  covering of the continuum, emission lines are not covered.}
\tablenotetext{d}{The spectra of these objects do not include
  H$\alpha$ in the bandpass.}
\tablenotetext{e}{These objects required two Gaussian components to
  model the Balmer absorption lines.  The FWHM measurements are for the
two-component feature, and the $\tau$-weighted velocity offset is
given.  }
\tablenotetext{f}{As noted by \citet{aoki10}, in this object the
  H$\alpha$ absorption line falls upon a prominent CO telluric
  feature, making the measurement of depth  uncertain.}
\end{deluxetable*}
\end{longrotatetable}

 The results are given in Table~\ref{specfit_tab}.  We report the FWHM of the
absorption feature and the optical depth (normalization).  For single
features, the uncertainties were taken from the {\tt Sherpa} models.
In the several cases where there were two components, we measured the
FWHM and optical depth normalization from the combined profile.  The 
uncertainties were generated by simulating 10000 profiles with
parameters drawn from normal distributions given by the {\tt Sherpa}
best fit and errors.  The 1$\sigma$ errors on the full profile were
obtained from the 
distribution of measurements from the simulated profiles.  We also
report the apparent hydrogen column density in $n=2$ from the
H$\alpha$ profile.  These  were computed in the usual way from
\citet[][Equation 9]{ss91}.  This can be regarded as a lower
limit to the column density, since it does not take into account 
partial covering.

\subsection{Partial Covering}\label{partcov}

It is well known that quasar absorption often exhibits partial
covering, i.e., 
the absorbing outflow does not completely occult the continuum
emission region \citep[e.g., see][for references and
  discussion]{leighly19}.    Partial covering is recognized in the
spectra presented in this paper through comparison of the optical
depths of the different 
Balmer lines.  Atomic physics requires that $\tau_1/\tau_2$ equals the
ratio of the products of the line wavelength and oscillator strength
\citep[e.g.,][]{ss91}.  In our case, the ratio of H$\beta$ to H$\alpha$
optical depths should be $0.14$, while the ratio of H$\gamma$ to
H$\beta$ should be $0.33$.  In nearly all cases, the best-fitting 
optical depths (Table~\ref{specfit_tab}) gave ratios larger than these
values, indicating the presence of partial covering, but not the
statistical necessity of partial covering.  Thus, our second set of
model fits constrained the relative optical depths to be the values
demanded by atomic physics.  We then compared the $\chi^2$ for the
constrained and unconstrained models using the F-test.
We found that
the spectra of 
six objects were consistent with no partial covering
(Table~\ref{specfit_tab}).  These objects had either very noisy
spectra or very low optical depth absorption
(Fig.~\ref{specfit_plots}.).   

The nature of partial covering is a complex subject; see
\citet{leighly19} for a discussion.  Given the fact that we are
constraining partial covering using only two or three lines, we used
step-function partial covering \citep[e.g.,][]{arav05} as the simplest
 model.  In this model, the absorber uniformly occults a
fraction of the continuum emission region given by $c_f$, while the
remainder is not occulted.  This model is complicated by the fact that
the continuum and the broad line region may have different covering
fractions \citep[e.g.,][]{ganguly99}.

Given the limits of the data, we adopt the following spectral-fitting
strategy to obtain an estimate of the true column density.  Since
partial covering is present, the apparent optical depth of H$\alpha$
yields only a lower limit of the column density of neutral hydrogen in
$n=2$.  As discussed above, partial covering is recognized in data
when the ratio of a 
weaker to stronger line is larger than predicted by atomic physics.
Therefore, the higher order lines yield higher  (i.e., better)
lower limits.  So we use the $F$ test to 
determine the smallest statistically necessary Balmer
line\footnote{For example, we compared the $\chi^2$ for a model with
H$\alpha$, H$\beta$, and H$\gamma$ with the $\chi^2$ for a model
including only H$\alpha$ and H$\beta$.  If the F test indicates that
H$\gamma$ 
is statistically necessary, it is the smallest statistically necessary
line, and the column density is determined from it.  If not, we repeat
the test with H$\beta$, and so on.}.  These are listed in
Table~\ref{specfit_tab}.  Another approach would have been to report
the optical depth from the best-fitting partial covering model.
However, given the range of signal-to-noise ratios exhibited by the
spectra, that result would have been model dependent.  Cleaving
closely to measurements obtained directly from the data provides a
more model-independent approach.

The covering fraction can be obtained directly from the optical depth
ratios \citep[e.g.][]{ganguly99, sabra01, arav05}, but in that case,
estimating the uncertainty in the covering fractions may not be
straightforward.  Instead, we use the flexibility of {\tt Sherpa} to
construct an analytical partial covering model.

The true column density is equal to the observed column density times
the covering fraction \citep[e.g.,][]{arav05}.  Also, in principle,
the continuum covering fraction and emission line covering fraction
can be measured separately \citep[e.g.,][]{ganguly99}.  But given the
SNR limitations of our spectra, we consider three limiting partial
covering scenarios.  In all of these, the higher order Balmer line
optical depths are constrained according to atomic physics, i.e.,
using the ratios given above.   
\begin{enumerate}
  \item {\bf P(C+L):} The absorber partially covers both the continuum
    and the emission line 
    region.  The true column density is the product of the covering
    fraction times the column density estimated from the smallest
    statistically necessary line.
  \item {\bf PC:} The absorber partially covers the continuum emitting region,
    leaving the emission-line region uncovered.    The true column
    density is the product of the covering fraction times the column
    density estimated from the smallest statistically necessary line.
  \item {\bf FC+PL:} The absorber completely covers the continuum emission region,
    and partially covers the emission-line region.  In this case,
    since the continuum is completely covered, the best estimate of
    the column density is that estimated from the smallest
    statistically necessary line (i.e., the covering fraction is 1.)
\end{enumerate}

We can make some inferences about the nature and location of the
absorber among the eight objects in which partial covering was
statistically necessary.  In six spectra, the H$\alpha$ line was too
deep to be consistent with the PC case, i.e., the continuum
partially covered and the emission lines uncovered. In fact, the two
objects that favored the PC model are the ones with no H$\alpha$
spectral coverage, and it is possible that the preferred spectral
model might be different if H$\alpha$ had been available.  This
preference 
suggests that  the absorber lies outside of or is cospatial with
the broad line region.
On the other hand, five were best fit by the FC+PL model, i.e., the
absorber completely covers the continuum, but only partially covered
the broad emission line region. This suggests that in most of the
objects, the absorber is located at a larger radius than the broad
line region, but it is not so distant that the BLR and continuum are
unresolved with respect to the unknown structure in absorbing gas that
is the origin of partial covering (i.e,. the case required by P(L+C)
partial covering).   

\begin{figure*}[!t]
\epsscale{1.0}
\begin{center}
\includegraphics[width=6.5truein]{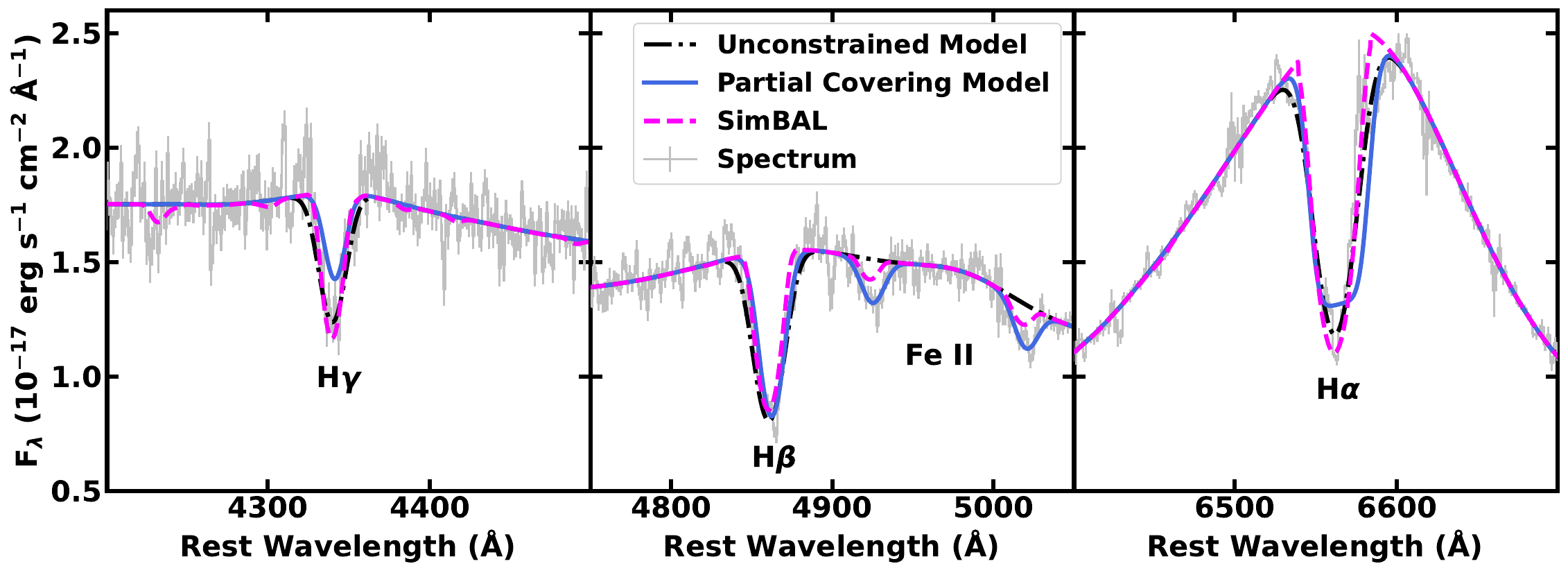}
\caption{An illustration of partial covering models in
  SDSS~1635$+$1439.  The black line shows the unconstrained model fit
  in which the relative Balmer optical depths were allowed to vary
  independently,   yielding an excellent fit but an unphysical model.  The
  blue line  shows the best-fitting step-function partial covering model.  The
  H$\alpha$ line is truncated because the absorption on the continuum
  is saturated and it is not black because the partially-covered BLR
  fills it in.  The magenta line shows an illustration of a {\tt
    SimBAL} model.  {\tt SimBAL} uses power-law partial covering
  \citep[e.g.,][]{leighly18, leighly19}, which allows the line to largely
retain its shape and yields an overall better fit.  We note that this
is not formal a {\tt SimBAL} model fit; those will be presented in
Choi et al.\ in prep.}    \label{simbal_example}
\end{center}
\end{figure*}

In three cases, the best-fitting partial covering model did not fit 
as well as the unconstrained model, i.e., the unphysical one in which all
absorption line optical depths are allowed to vary independently.  The
best-fitting unconstrained and partial covering models for
SDSS~J1635$+$1439 are shown in Fig.~\ref{simbal_example}.  The Balmer
absorption apparent optical 
depth in this object is the largest in the sample, but the simple
step-function partial covering mdoel is not consistent with the shape
of the H$\alpha$ absorption line.  A more complex model such as the
power-law partial covering model \citep[e.g.,][]{dekool02c,   sabra05,
  arav05},  in which $\tau(x)=\tau_{max} x^a$ where $x$ is the
fractional surface area, may be justified. Fig.~\ref{simbal_example}
shows an example of a {\tt SimBAL} model\footnote{We note that this is
not a formal {\tt SimBAL} model fit to the broadband spectrum, but
rather a by-hand $\chi$-by-eye model fit; the formal {\tt SimBAL}
model fit will be included in Choi et al.\ (in prep.).} overlaid on
the partial covering model for the purpose of illustration. The
power-law partial covering model both fits the line ratios better
(especially H$\gamma$), and it also retains the shape of H$\alpha$
better.  

Along with the optical depth measurements (i.e., the maximum optical
depth, the apparent column density of H$\alpha$, and the true column
density estimate that takes into account the partial covering) given
in Table~\ref{specfit_tab}, we computed three measures of $dN/dv$,
i.e., the optical depth or column density divided by the line FWHM,
yielding an approximate measure of the opacity per speed
interval. This parameter is motivated if the acceleration mechanism is
radiative line driving, as the amount of radiation absorbed depends on
there being sufficient ions present but also that there be a velocity
dispersion so that an ion absorbs a wide band of continuum
photons rather than the relatively few continuum photons absorbed if
the gas is stationary \citep{sobolev57}.  The complex BAL spectra which 
includes absorption from many excited states ensures that sufficient
ions are present for the gas to be accelerated by radiative line
driving (the $dN$ part), while the width of the line indicates how
much of the continuum will be absorbed (the $dV$ part).  

\subsection{Accretion Properties}\label{otherprops}

We followed nearly the same procedure as \citet{leighly22} to
estimate the bolometric luminosity, black hole mass, and Eddington
ratio, as described below. The values are given in
Table~\ref{global_tab}.  

To estimate the bolometric luminosity and black hole mass, we 
needed measurements of the rest-frame flux density at 5100\AA\/ and
3$\mu \mathrm{m}$.  We used a linear interpolation of the log of the
flux densities of the nearest larger and smaller wavelength photometry
points.  To estimate the uncertainty, we repeated the interpolation
10,000 times, perturbing the flux density values using a normal
distribution with standard deviation equal to the photometry errors.   

As in \citet{leighly22}, we estimated the bolometric 
luminosity using  the rest-frame flux density at 3$\mu \mathrm{m}$ and
the bolometric correction derived by \citet{gallagher07}.  
The black hole mass requires an estimate of the radius of the
H$\beta$-emitting broad-line region.  As in \citet{leighly22}, we used
the formalism given by \citet{bentz06}.  A difference between that
paper and this one is that we estimated the luminosity density at
5100\AA\/ from the photometry rather than the continuum measurement
from spectral fitting because our near-IR spectra are not 
reliably fluxed.  We did not correct for reddening intrinsic to the
quasar, noting that the photometry SEDs show that most appear
unreddened.  The black hole mass was estimated using the FWHM of the
H$\beta$ line and following the formalism of \citet{collin06}.  In
particular, we used their Equation 7 to estimate the scale factor $f$
based on the FWHM of the H$\beta$ line.

$\alpha_{oi}$, used previously by \citet{leighly24} to characterize
the optical--NIR spectral energy distribution, is a measure of both
the reddening (the blue half) and the relative strength of the upturn
longward of the 1$\mu\mathrm{m}$ dust-sublimation break that is
thought to be emission from the hot dust on the inner edge of the
torus (the red half)\footnote{In \citet{leighly24}, we also estimated
$D_\mathrm{red}$ and $D_\mathrm{torus}$, measures of the difference
between the mean SED from the unabsorbed comparison sample introduced
in \citet{leighly22} on the red side and blue side, respectively.
These two parameters break the degeneracy between reddening and torus
strength in
$\alpha_{oi}$ \citep[Table 1, Fig.\ 3 in][for definitions and
  discussion]{leighly24}.  The photometry data used in this paper are 
not robust enough to estimate $D_\mathrm{red}$ and $D_\mathrm{torus}$,
chiefly from the lack of  observed-frame near-infrared (rest-frame
optical) photometry in 7 objects, leaving some ambiguity in how to
interprete the $\alpha_{oi}$ results presented here.}.  For reference,
the \citet{krawczyk13} quasar composite spectrum yields  $\alpha_{oi}=
-0.97$. 

The bolometric luminosity $L_\mathrm{Bol}$, the black hole mass, the
Eddington ratio, and $\alpha_{oi}$ are given in
Table~\ref{global_tab}.  As before, the uncertainties were generated
from the distribution of 10,000 random draws of measurements from
perturbed flux values.
  
\begin{deluxetable*}{lCCCC}
\tabletypesize{\scriptsize}
\tablecaption{Accretion Properties \label{global_tab}}
\tablehead{
  \colhead{Object Name} & \colhead{$\log$ Bolometric
    Luminosity\tablenotemark{a}} & 
  \colhead{$\log$ Black Hole Mass\tablenotemark{a}} &
  \colhead{$\log  L_\mathrm{Bol}/L_\mathrm{Edd}$\tablenotemark{a}}  &
  \colhead{$\alpha_{oi}$} \\ 
\colhead{} & \colhead{[erg\, s$^{-1}$]} & \colhead{[solar masses]} &
\colhead{} & \colhead{} \\
}
\startdata
080202.69+140315.1 & 46.99 \pm 0.04 & 9.4 \pm 0.05 & -0.52\pm 0.06 &
-1.43\pm 0.13 \\
083942.11+380526.4 & 46.95 \pm 0.05 & 9.3 \pm 0.05 & -1.45 \pm 0.06 &
-1.52 \pm -0.14 \\
085910.40+423911.3 & 46.76 \pm 0.02 & 9.11^{+0.01}_{-0.008} & -0.45 \pm
0.02 & -1.15 \pm 0.03 \\
101927.37+022521.3 & 47.01 \pm 0.015 & 9.16^{+0.02}_{-0.009} & -0.25 \pm
0.02 & -0.93 \pm 0.03 \\
112526.12+002901.3 & 46.29\pm 0.02 & 8.9\pm 0.025 &
-0.71^{+0.009}_{-0.05} & -1.24 \pm 0.03 \\
122933.32+583427.6 & 47.37 \pm 0.04 & 9.56\pm 0.03 & -0.29 \pm 0.05 &
-1.24 \pm 0.08 \\
124452.49+583427.6 & 47.05 \pm 0.025 & 9.19 \pm 0.008 & -0.24 \pm 0.03
& -1.00 \pm 0.04 \\
143916.28+162858.5 & 47.14 \pm 0.02 & 9.27 \pm 0.015 & -0.23 \pm 0.03 &
-0.92 \pm 0.04 \\
160915.16+561943.2 & 46.45 \pm 0.02 & 8.96 \pm 0.009 & -0.61 \pm 0.02 &
-1.23 \pm 0.03 \\
162119.22+081950.7 & 47.47 \pm 0.03 & 9.22 \pm 0.01 & 0.14 \pm 0.03 &
-0.59 \pm 0.05 \\
162527.73+093332.8 & 47.26 \pm 0.03 & 9.35 \pm 0.06 & -0.19 \pm 0.06 &
-1.27 \pm 0.14 \\
163515.87+143925.9 & 46.53 \pm 0.02 & 9.22 \pm 0.03 & -0.79 \pm 0.04 &
-0.94 \pm 0.04 \\
164419.75+530750.4 & 46.39 \pm 0.02 & 9.06 \pm 0.09 &
-0.77^{+0.09}_{-0.08} & -1.06 \pm 0.05 \\
172341.08+555340.5 & 47.41 \pm 0.02 & 9.31 \pm 0.02 & 0.00\pm 0.03 &
-1.28 \pm 0.06 \\
\enddata
\tablenotetext{a}{Parameter uncertainies are estimated using
  measurement errors (e.g., uncertainties in flux density or FWHM)
  only.   Systematic errors in single epoch estimates of black
  hole masses are typically a few tenths dex \citep[e.g.,][]{vp06}.} 
\end{deluxetable*}

\section{Correlations}\label{correlations}

\begin{figure*}[!t]
\epsscale{1.0}
\begin{center}
\includegraphics[width=6.5truein]{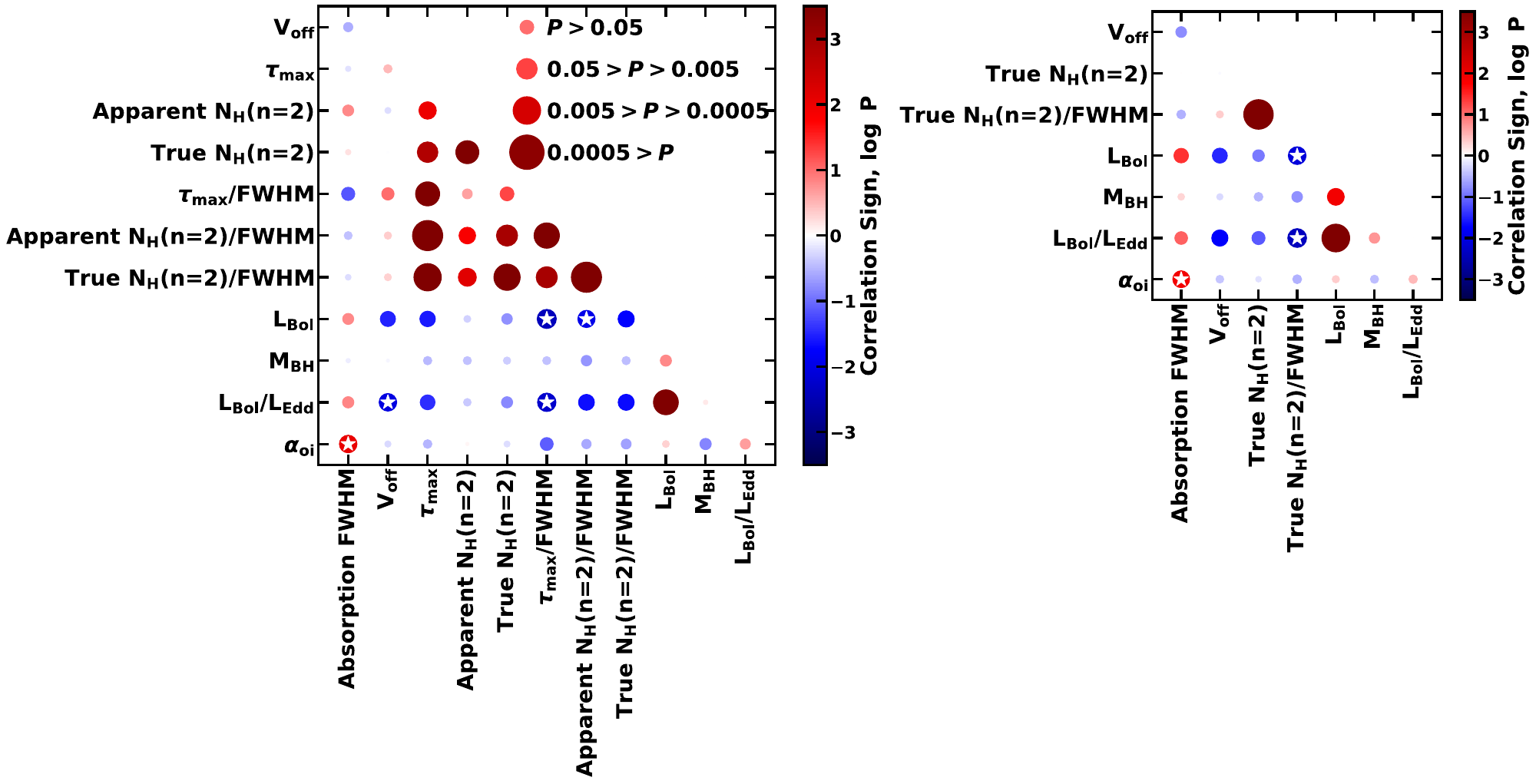}
\caption{The results of the Spearman rank correlation analysis for the
  12-object subsample that covers H$\alpha$ (left), and for the full
  14-object sample (right).  The size  and  color of each  marker indicate
  the sign and $p$ value of the correlation. Anticorrelations are shown
  in blue, and correlations are shown in red. The shade of the color
  of each point indicates the significance of the correlation as a
  continuous variable, while the discrete sizes of the points
  characterize a range of $p$ values: $p > 0.05$, $0.05 > p > 0.005$,
  $0.005   > p > 0.0005$, and $0.0005 > p$. The circular markers show
  the results   for the full sample.  Most of the   strongest positive
  correlations are   trivial, including those between   measures of
  opacity, and   $L_\mathrm{bol}$ versus the Eddington ratio. The
  stars  mark the   physically interesting correlations that survive
  the one-out   criterion (see text for  details).  
   \label{correlation_plots}}
\end{center}
\end{figure*}

In order to gain insight  into the nature of Balmer absorption in
FeLoBAL quasars, we performed a correlation analysis.   The Spearman
rank correlation is appropriate for these data, since there is no
reason to expect that the measurements should have a Gaussian
distribution.   We
represent the results in Fig.~\ref{correlation_plots} in the same way
as in \citet{leighly22} and \citet{leighly24}.  Specifically, the plots
represent the log of the $p$-value for the correlation, where the sign
of the value represents the sense of the correlation.  For example, a
large negative value implies a highly significant anticorrelation.  
We computed correlations separately for the sample of 12 objects that
all have H$\alpha$ in their spectra separately from the 14 object
sample that includes two objects that only have H$\beta$ in their
spectra.  

As discussed in \S~\ref{analysis}, the column density estimates are
uncertain 
because although there is evidence for partial covering in all
objects, the amount and nature of the partial covering is difficult to
determine.  When partial covering is not taken into account, the
column density measurements are a lower limit.  In principle, this
means that we should use correlation methodology that takes into
account lower limits.  However, those methods are generally intended
for cases when all the data is uniform and the limits arise from
detectability, e.g., as in a flux-limited sample.  This is not the
case for our spectra; in a sense, all of our measurements are limits.
Therefore, we do not explicitly take lower limits into account but
rather bear these limitations in mind in the analysis.  We 
also plan compare these results with the {\tt SimBAL} modeling results
which  accounts for the partial covering explicitly using the full
spectrum (Choi  et al.\ in prep., Leighly et al.\ in prep.)

Fig.~\ref{correlation_plots} shows that there are a number of
correlations with $p$ values less than our cutoff of 0.05.  Many 
of these are trivial; for example, all measures of optical depth and
column density are correlated with one another, and $L_\mathrm{Bol}$
is correlated with $L_\mathrm{Bol}$/L$_\mathrm{Edd}$. Because the
sample is small, it is  possible that some of these correlations are
accidental.  We therefore counted a correlation as significant only if
correlations of all combinations of samples obtained by taking one
object out satisfy our $p<0.05$ significance criterion (hence referred
to as the one-out criterion). These correlations are marked with white
stars in Fig.~\ref{correlation_plots} and discussed below.   

\subsection{Correlation between $L_\mathrm{Bol}$
  and Eddington Ratio with $V_\mathrm{off}$} \label{eddrat_corr}

\begin{figure*}[!t]
\epsscale{1.0}
\begin{center}
\includegraphics[width=4.5truein]{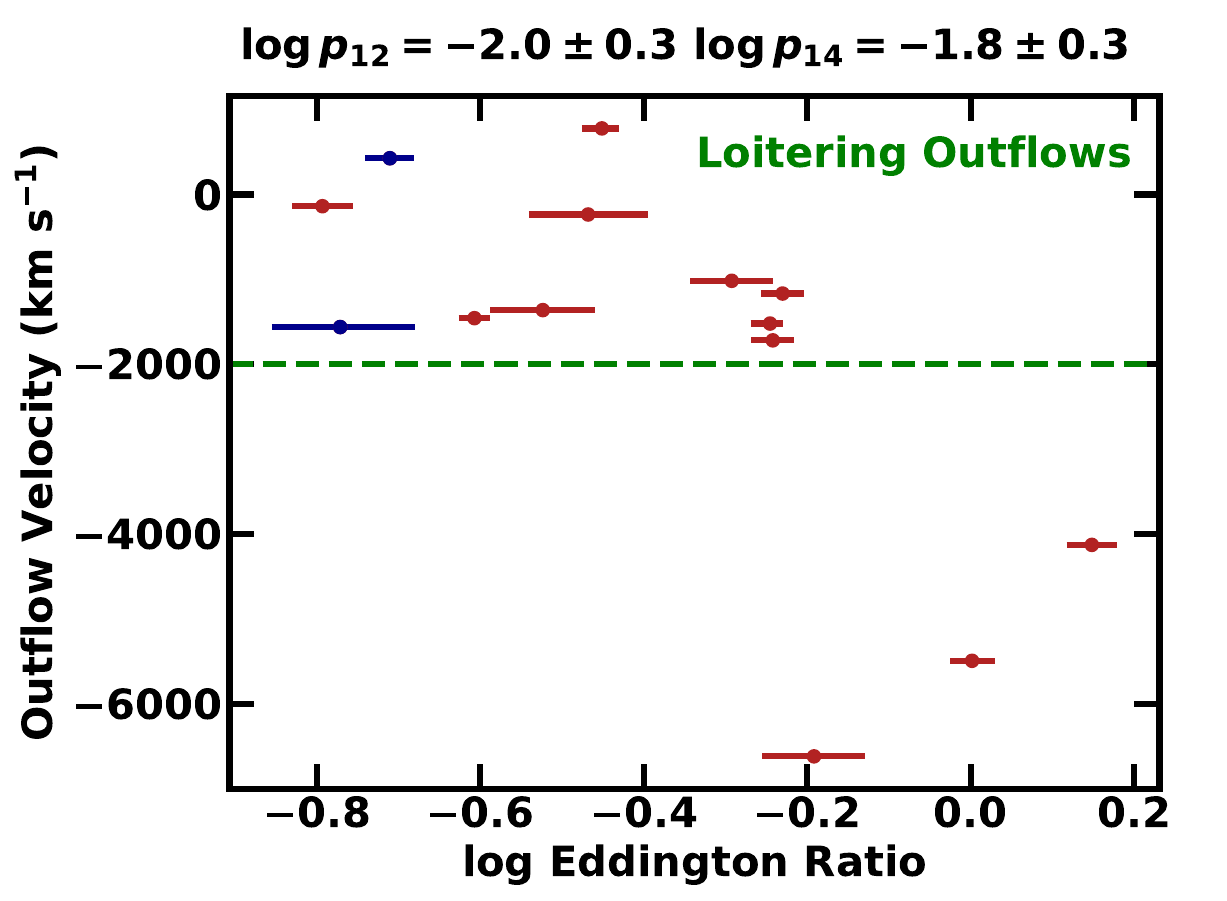}
\caption{The anticorrelation between Eddington ratio and velocity
  offset. The red points mark objects that have H$\alpha$ in the 
  bandpass; the velocity offset is measured from H$\beta$ for the two
  objects with blue points.  The $p$ values included above the plot
  were computed for   the 12-member sample that includes H$\alpha$ in
  the spectrum   ($p_\mathrm{12}$) and the full sample
  ($p_\mathrm{14}$). The   anticorrelation is common among BAL quasars but is weak in 
  this sample, most likely because most of the targets belong to the
  loitering outflow class (above the green dashed line).  First
  identified by \citet{choi22},   loitering outflow quasars are
  identified by their low outflow speeds   and compact outflows.
   \label{velcor_plot}}
\end{center}
\end{figure*}

We found a significant anticorrelation between the velocity offset and 
Eddington ratio for the 12 object sample (Fig.~\ref{velcor_plot}).
The 14 object sample just misses our cutoff with three of the one-out
correlations having $p=0.055$.  The anticorrelation between
$V_\mathrm{off}$ and $L_\mathrm{Bol}$ did not survive the 
one-out criterion imposed above, but because $L_\mathrm{Bol}$ and
Eddington ratio are strongly correlated, it might be safe to assume
that this correlation is present as well. 

The anticorrelation between $V_\mathrm{off}$ and $L_\mathrm{Bol}$ is
 ubiquitous among BAL quasars.   \citet{laor02} and
\citet{ganguly07} found that the behavior is more like an upper limit
envelope so that a quasar needs to have a certain luminosity to
attain a very large outflow velocity, but lower velocities are also
observed.  \citet{choi22} found a correlation between $L_\mathrm{Bol}$
and velocity offset among low-redshift FeLoBAL quasars.

In our case, the anticorrelation is not very strong, and examination
of Fig.~\ref{velcor_plot} shows why.  Eleven of of the fourteen
objects that show Balmer absorption are loitering outflow FeLoBAL
quasars.  This class of objects was first identified by \citet[][see
  their   Fig.\ 18]{choi22}.  Specifically, \citet{choi22} found that
most of the low-redshift FeLoBAL quasars show an
anticorrelation between the outflow location and the outflow speed
such that outflows located close to the central engine have the
highest outflow speeds.  (This is the behavior expected if the
outflows are accelerated by radiative line driving, because the
radiation flux is largest close to the central engine.)  The loitering
outflow FeLoBALQs are were identified as outliers from this
relationship with $\log R < 1$ [pc] and $|V_\mathrm{off}| < 2000\rm \,
km\, s^{-1}$, i.e., very low outflow speeds for their location close
to the central engine.  While we do not yet have the {\tt SimBAL}
measurements of $\log R$ for the high-redshift FeLoBAL sample, and we  
expect that the criterion will scale with luminosity (i.e., the
criterion may be better expressed as in terms of $\log
R/R_\mathrm{sub}$, where $R_\mathrm{sub}$ is the dust sublimation
radius), we can confirm that nearly all of the objects in our sample
have the \ion{Fe}{2} absorption morphology typical of a loitering
outflow; specifically, strong absorption from highly-excited states.
Thus the correlation seen in this sample is essentially between the
the loitering outflow candidates, which have $V_\mathrm{off}<2000
\rm km\, s^{-1}$, and the three objects with larger velocities:
SDSS~J1621$+$0819, SDSS~J1625$+$0933, and SDSS~J1723$+$5553.  Of these
three, only SDSS~J1723$+$5553 has \ion{Fe}{2} absorption morphology
more typical of an ordinary FeLoBALQ; the other two show the strong
highly-excited-state \ion{Fe}{2} absorption characteristic of
loitering BAL outflows.  In other words, we
cannot expect strong correlations with outflow velocity in a sample
dominated by a type of object that is characterized by a limited range
of outflow velocities.

We note that the fact that nearly all of these objects are loitering
outflow FeLoBALQs is consistent with the result from \S\ref{partcov}
that most of the objects in which partial covering could be detected are
best fit by the FC+PL model.  {\tt SimBAL} model fitting results of
loitering outflow FeLoBALQs generally find a location near the torus,
consistent with the partial covering results, i.e., close enough that
the angular size of the broadline region is resolved, but the angular
size of the continuum emitting region is not  \citep[for further
  discussion, see][]{leighly19, choi22b}.

\subsection{Correlations between $dN/dv$ Measures and Eddington
  Ratio} \label{dndvcorr}

\begin{figure*}[!t]
\epsscale{1.0}
\begin{center}
\includegraphics[width=6.5truein]{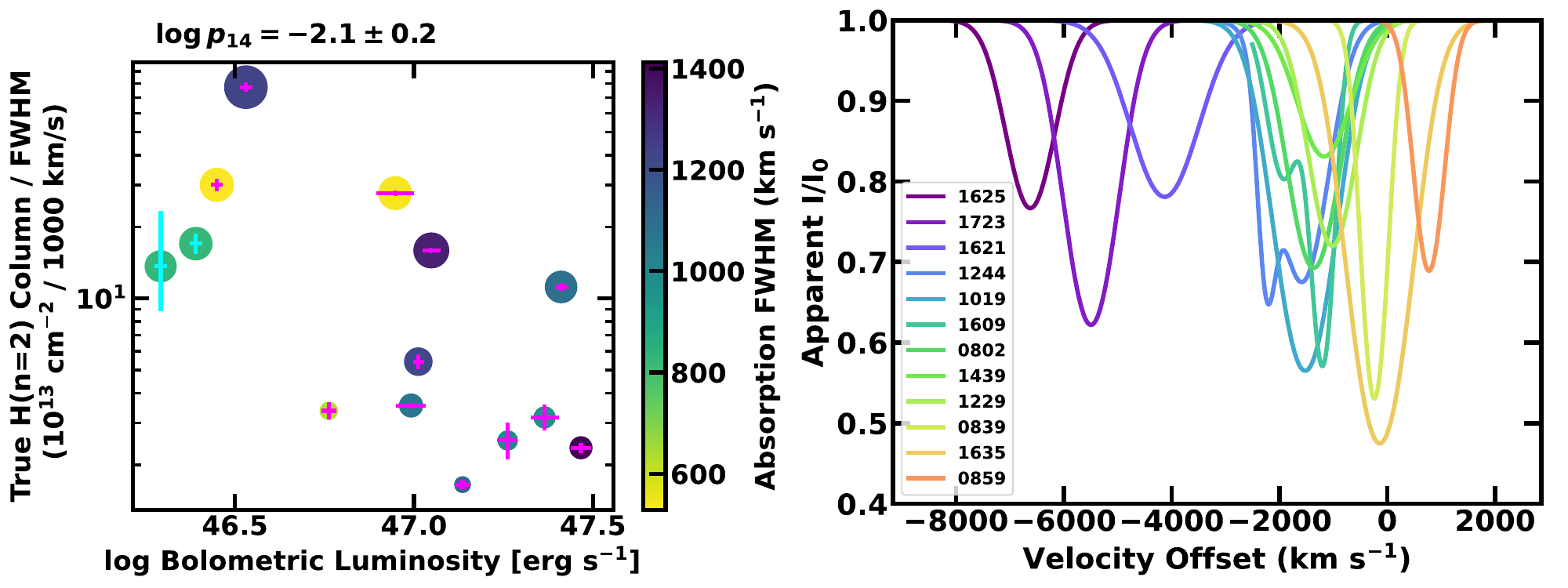}
\caption{{\it Left:} An example of one of the anticorrelations between
  measures of $dN/dv$ and   $L_\mathrm{Bol}$ or Eddington ratio.  The
  log of the significance   for this correlation is given at the top
  of the plot.  The circular marker color indicates the absorption
  line FWHM, which is weakly correlated with $L_\mathrm{Bol}$,
  while the circular marker size indicates the log of the true column
  density, which is weakly anticorrelated with $L_\mathrm{Bol}$.  The
  errors on the parameters are shown in pink and cyan for the two
  objects that lack H$\alpha$ observations. {\it Right:} 
  The twelve H$\alpha$ absorption profiles labeled by their RAs in
  inverse order of outflow speed.   There is a general trend for
  higher velocity lines to be shallower and possibly broader.} 
   \label{dndvcorr_plot}
\end{center}
\end{figure*}

We found that $L_\mathrm{Bol}$ and the Eddington ratio are
anticorrelated with several measures of the differential line opacity
such that higher values of $dN/dv$ are found in objects with lower
outflow velocity widths. 
Five such correlations among the 12 and 14 object samples survive the
one-out correlation criterion; Fig.~\ref{dndvcorr_plot} shows one of
them. Fig.~\ref{dndvcorr_plot} also  shows the H$\alpha$ profiles for
the sample.  Despite the fact that 
this plot only shows the apparent optical depth profile and partial
covering is not taken into account, it gives the general impression
that higher velocity outflows tend to be both shallower and broader.

Among the interesting correlations marked with stars in
Fig.~\ref{correlation_plots}, we found stronger correlations among the
$dN/dv$ measures than the integrated opacity measures.  Because $dN/dv$
is a ratio, we can ask whether the anticorrelation arises from 
an anticorrelation between column density and $L_\mathrm{Bol}$, and/or a
correlation between FWHM and $L_\mathrm{Bol}$ (likewise between these
two parameters and Eddington ratio).  As shown in
Fig.~\ref{correlation_plots}, both relationships are present, although
they are weak; only FWHM versus $L_\mathrm{Bol}$ satisfies our $p<0.05$
criterion, while the others range from $0.085 < p < 0.18$.   The
dependence on these two parameters is illustrated in
Fig.~\ref{dndvcorr_plot} by the colorbar and symbol size, respectively.

\subsection{Correlation between $\alpha_{oi}$ and Absorption Line
  FWHM} \label{alphaoi_corr} 

\begin{figure*}[!t]
\epsscale{1.0}
\begin{center}
\includegraphics[width=4.5truein]{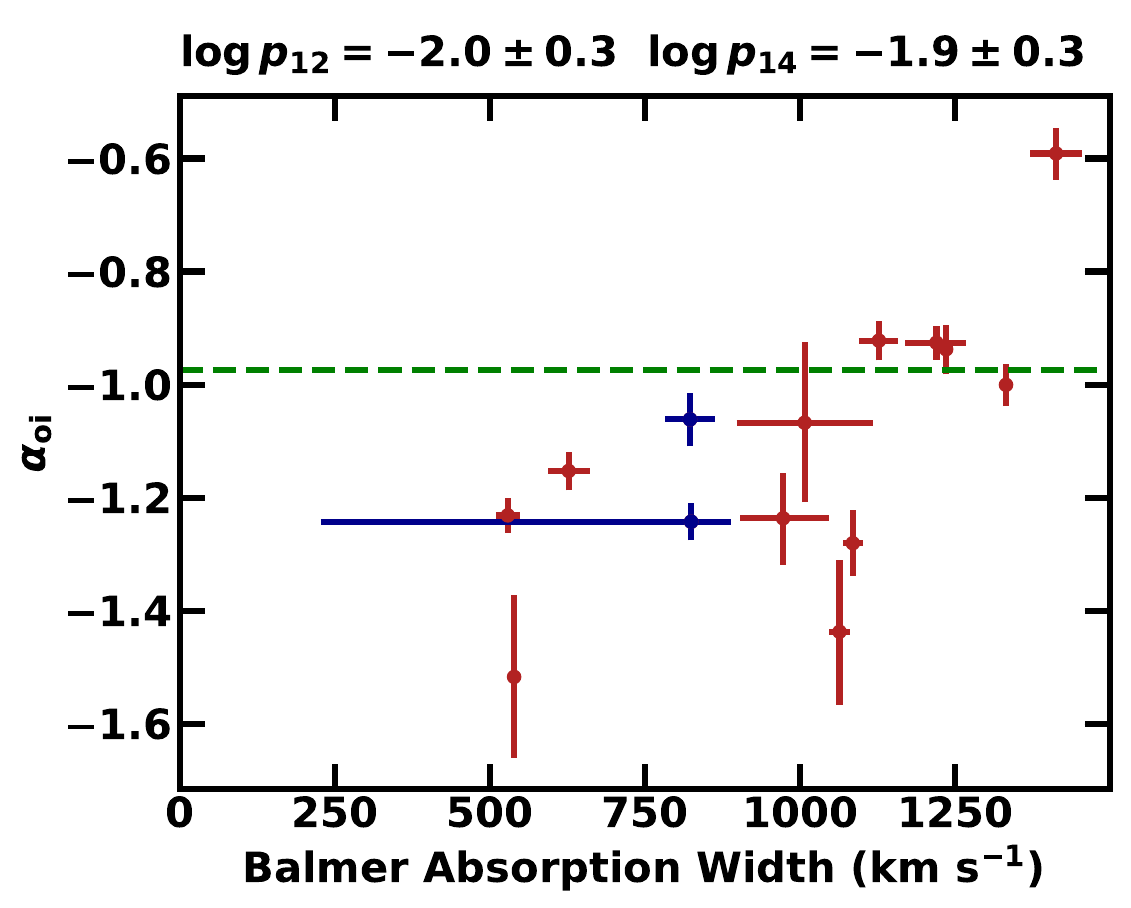}
\caption{Relationship between the absorption line width and
  $\alpha_\mathrm{oi}$, the point-to-point slope between 5100\AA\/ and
  3$\mu \mathrm{m}$.  The red points mark objects that have H$\alpha$ in the 
  bandpass; the velocity offset is measured from H$\beta$ for the two
  objects with blue points.   The horizontal line marks the $\alpha_{oi}$ value
from the \citet{krawczyk13} composite spectrum.  Objects with values
near or below this line have no reddening and weak torus
emission.   The object with the large value of $\alpha_{oi}$ is
SDSS~1621$+$0819, and the value is large due to a large flux in {\it
  WISE} W3, with signal-to-noise ratio of 13.9. }
   \label{fwhm_vs_alphaoi_plot}
\end{center}
\end{figure*}

The final correlation that survived the one-out criterion is the 
positive correlation between $\alpha_{oi}$ and absorption line FWHM
shown in Fig.~\ref{fwhm_vs_alphaoi_plot}.  

$\alpha_{oi}$ was a key parameter in the analysis of low-redshift
FeLoBAL quasars presented by \citet{leighly24}.   As noted above,
$\alpha_{oi}$ simultaneously measures reddening and the strength of
the torus \citep[see Fig.\ 3 in][]{leighly24}, with larger (i.e., more
positve and flatter) values indicating stronger reddening and stronger 
torus emission.  In that analysis, $\alpha_{oi}$ was strongly
anticorrelated with the location of the outflow, but among the
unabsorbed comparison sample, it was also strongly correlated with the
Eddington ratio and black hole mass \citep[see Fig.\ 11
  in][]{leighly24} such that objects with higher accretion rates
showed large (more positive and therefore flatter) values of
$\alpha_{oi}$. Because the unabsorbed comparison objects showed little
evidence for reddening \citep[Fig.\ 4 in][]{leighly24}, these larger
values correspond stronger hot dust emission. $\alpha_{oi}$ was also 
correlated with $R_\mathrm{FeII}$, the ratio of the optical
\ion{Fe}{2} emission to the H$\beta$ emission.  This ratio is broadly
thought to be related to quasar fundamental physical parameters such
as black hole mass, accretion rate, and orientation
\citep[e.g.,][]{shen_ho_14}.  Thus, it could be that the correlation
observed in this sample is a side effect of the anticorrelation
between Eddington ratio and outflow velocity discussed above.

It is notable that in this sample, nearly all objects have
$\alpha_{oi}$ values steeper than the quasar SED composite spectrum
from \citet{krawczyk13}.  This behavior was also found among
  the low-redshift loitering outflows \citep[][Fig.\ 13]{leighly24}.
Specifically, the SEDs of those objects (with nore negative
$\alpha_{oi}$ values) were consistent with no reddening and notably
weak torus emission.  In \citet{leighly24} we linked the weak torus
emission to the low accretion rate typical of loitering outflow
objects (\S 8.1) of that paper.  The lack of reddening among the
loitering outflow FeLoBAL quasars contrasts with the general trend
that BALQs tend to be reddened compared with unabsorbed quasars
\citep[e.g.,][]{krawczyk15}. 

\section{Discussion}  \label{discussion}

\subsection{Population Properties}  \label{properties}

Balmer absorption is rare among FeLoBAL quasars, and even rarer among
the general population of quasars.  Studying the properties of extreme
objects can give us valuable insight into the phenomenon in general;
such a study is even more valuable when a sample of such objects can
be studied.  Previously, the literature on Balmer absorption consisted
of reports of the phenomenon in one or two objects.  Given the rarity
of examples of Balmer absorption lines observed in quasar spectra,
this sample of 14 objects is a significant addition to the literature
and reveals important insights into this class of quasar outflows.

Nearly all of the objects with H$\beta$ absorption lines are
classified as loitering outflow FeLoBALQs.  As outlined above,
\citet{choi22} 
developed this class to describe objects with compact, low-velocity
outflows that were outliers from the radius and outflow speed 
anticorrelation exhibited by most of the FeLoBAL quasars
\citep[][Fig.\ 18]{choi22}. \citet{leighly22} found that low-redshift
FeLoBAL quasars were characterized by either a low accretion rate or a
high accretion rate, while intermediate accretion rates more typical
of an unabsorbed quasars were not observed.  The loitering outflow
quasars in those papers were all members of the low accretion rate
group, which had a median Eddington ratio  of ($\log
L_\mathrm{Bol}/L_\mathrm{Edd}=-0.95$).  The 
median Eddington ratio of this sample is higher
($\log L_\mathrm{Bol}/L_\mathrm{Edd}=-0.37$). However, that difference
is     likely to be a consequence of the flux limited nature of SDSS,
and observational feasibility.  Low-accretion-rate
objects produce less light and are predicted to fall out of
flux-limited samples at higher redshifts \citep{jester05}, which means
that high-redshift loitering outflow FeLoBALQs are even rarer in SDSS
than are the lower-redshift ones.  In addition, they needed to be
bright enough to obtain NIR spectra of sufficient SNR on $<8$-meter
class telescopes in a reasonable amount of time. 

Another hallmark of a low accretion rate is broad Balmer emission
lines, strong [\ion{O}{3}], and weak \ion{Fe}{2}
\citep[e.g.,][]{shen_ho_14}. \citet{leighly22} found these signatures
among the low accretion rate FeLoBAL quasars (Fig.\ 9 in that paper).
While some of the spectra  shown in Fig.~\ref{specfit_plots} have
strong [\ion{O}{3}] \citep[e.g., SDSS~0859$+$4239, SDSS~1019$+$0225;
  see also][]{schulze18}, others do not.  However, [\ion{O}{3}] is
known to show a particularly strong Baldwin effect \citep[e.g.,][]{netzer04,
  stern12, wang25}, so we can expect weaker [\ion{O}{3}] emission
lines in a higher-redshift (and therefore higher luminosity) sample.
Whether or not there are detectable differences among the optical
emission lines in our high-redshift FeLoBALQ sample will have to wait
for analysis of the remainder of the sample and the unabsorbed comparison
sample (Leighly et al.\ in prep.).  

As mentioned above, \citet{choi22} used {\tt  SimBAL} to extrapolate
the best spectral models for our sample of low-redshift FeLoBAL
quasars to the rest-frame optical band to predict the presence of
Balmer absorption (Fig.\ 16 in that paper).  They found that {\it all} 
compact outflows ($\log R < \sim 1$ [pc]) predicted significant
opacity in Balmer absorption lines.  This included the loitering
outflow FeLoBALQs, but also the overlapping trough quasars. 
\citet{hall02} described this class of absorption lines:
``the troughs remain deep at velocities comparable to the spacing of
absorption troughs from different transitions, so that there are no
continuum windows between the troughs.'' {\tt SimBAL} analysis shows
that overlapping FeLoBALQs have the same high-excitation \ion{Fe}{2}
as the loitering outflows; the difference is that the lines are very
broad ($> 4000\, \rm km\, s^{-1}$ versus $< 2000\, \rm km\, s^{-1}$ for the 
loitering outflows) so that individual transitions are difficult to
identify.  Both loitering outflow and overlapping trough FeLoBALQs
have outflows located close to the central engine, but they are
strikingly different in other ways.  While loitering outflows are
characterized by a low accretion rate, overlapping trough quasars are
characterized by a high accretion rate \citep{leighly22}.  While the
speeds of low accretion rate objects increase with radius, the
speeds of high accretion rate objects decrease with radius
\citep{choi22b}.  Other differences in the structure of the outflows
including the volume filling factor are discussed by \citet{choi22b};
specifically, the log volume filling factor was foudn to be between
$-6$ and $-4$ in most objects, but was as high as $-1$ for the
loitering outflow FeLoBALQs.
Our full sample of near-IR spectra from more than 30 FeLoBALQs includes
overlapping trough objects, and our preliminary analysis does not
reveal 
any obvious Balmer absorption.  One reason that we do not identify
Balmer absorption in those spectra that it might be very broad and
shallow so that it blends into the continuum.   Thus, the sample
presented in this paper do not have overlapping trough FeLoBALQs by
selection.   

\subsection{Implications for BAL Acceleration  Mechanisms}  \label{acceleration }

Among the Balmer lines, we found evidence that $dN/dv$ is
significantly anticorrelated with Bolometric luminosity and Eddington
ratio. This correlation is a consequence of weaker correlations 
between optical depth (negative) and FWHM (positive) with
$L_\mathrm{Bol}$ and Eddington ratio.  We also found that the outflow
velocity is correlated with the Eddington ratio.

While the correlation between outflow velocity and bolometric
luminosity is a well-known property of BAL quasars
\citep[e.g.][]{laor02, ganguly07}, $dN/dv$ is not a parameter that is
usually computed, so it is difficult to compare with other studies.
However, we observe in our sample that the lines are
shallower at higher velocities (Fig.~\ref{dndvcorr_plot}), and that is
a well-known property of BAL outflows in general.  For example, the
composite spectra discussed by \citet{hamann19} clearly illustrate
this property, but that is only one of many such studies \citep[see
  also][]{rankine20}.  Although 
these studies are usually 
performed on \ion{C}{4}, which is generally saturated so that the
column density cannot be estimated from the depth of the  trough, the
behavior is roughly consistent with expectations from radiative line
driving, i.e., it requires much more momentum and energy to accelerate
gas to higher velocities.

We can use our Balmer column  measurements to roughly examine the
behavior of the mass outflow rate and similar properties.  The mass
outflow rate 
requires an estimate of the location, the total column density, and
the outflow velocity.  Because nearly all the objects are loitering
outflows, we can assume to zeroth order that they all have nearly the
same outflow location.  Because Balmer absorption requires extreme
physical conditions (i.e., high column density), we can assume that
the H(n=2) column density is proportional to the total column density
of the outflow.  This means that the mass outflow rate, $\dot
M_\mathrm{out}$, should be proportional to the product of the outflow
speed $|V_\mathrm{off}|$ and the H(n=2) column density.  Likewise, we
can compute a momentum flux measure as $\dot p \propto |V_\mathrm{off}|
\dot M_\mathrm{out}$, and a kinetic luminosity measure
$L_\mathrm{KE} \propto |V_\mathrm{off}| \dot p$. 

We investigated whether these measures  are
correlated with $L_\mathrm{Bol}$ and the Eddington ratio.  It turns
out that there is no correlation between our $\dot M$ measure and
either  $L_\mathrm{Bol}$ or the Eddington ratio for either the
12-object or the 14-object  sample;
in each case, $p \sim 0.5$.  This behavior arises because
$L_\mathrm{Bol}$ and the Eddington ratio are weakly correlated with
the outflow speed (i.e., taken to be positive for an outflow), and
weakly anticorrelated with the true N$_\mathrm{H}$(n=2) column
density.  This behavior contrasts with the general behavior of quasar
outflows, which show a strong correlation between the $L_\mathrm{Bol}$
and $\dot M_\mathrm{out}$ \citep[e.g.,][their Fig.\ 21]{choi22}.
Physically, our result might suggest that $\dot M_\mathrm{out}$ is
conserved, and is accelerated to higher velocities when the luminosity
and Eddington ratio are increased.  However, our $\dot M_\mathrm{out}$
measures span a large range in the sample (a factor of 14), while
$L_\mathrm{Bol}$ and Eddington ratio measurements span only one order
of magnitude, so such an interpretation would be overly simplistic. 
We plan to compare these inferences with the more precise {\tt SimBAL} 
measurements (Choi et al.\ in prep.).

In contrast, since $\dot p$ and $L_\mathrm{KE}$ include additional
factors of $|V_\mathrm{off}|$, they are both largely correlated with
both $L_\mathrm{Bol}$ and Eddington ratio ($0.015 < p < 0.059$) in 
agreement with previous studies \citep[e.g.,][and references
  therein]{choi22}.  

\subsection{A Comparison with the Properties of Little Red Dots}  \label{LRD}

One of the reasons that we decided to present the Balmer absorption
line properties in this subsample of near-IR spectra from 
our high-redshift FeLoBAL quasar sample is because
of the recent discovery of this absorption line in the so-called
Little Red Dots (LRDs).  While the precise classification of LRDs is
under some debate, they are generally taken to be a commonly-found
object at $z>4$ inferred to be an active galaxy from their
broad Balmer emission lines \citep[e.g.,][]{kocevski23}, although
alternative explanations for the broad lines have recently surfaced
\citep[e.g.,][]{baggen24}.  Typically, their SED 
has a V-shape \citep[e.g.,][]{kocevski25}.  In this section, we
compare the properties of Little Red Dots with FeLoBAL quasars, in
particular with the loitering outflow FeLoBAL quasars. 

\subsubsection{Balmer Absorption Lines} \label{balmer}

Balmer absorption lines have been found a number of LRDs
\citep{deugenio25, kocevski25, matthee24}.  The fraction of LRDs with
Balmer absorption has been estimated to be as high as 20\%
\citep[e.g.,][]{inayoshi25}.  The Balmer absorption appears in
H$\alpha$ and sometimes H$\beta$.  It is generally narrow (FWHM$\sim
200\rm\, km\, s^{-1}$) and mildly blueshifted.

It is dangerous to assume an observation phenomenon (Balmer absorption
lines) implies a common origin.  However, the properties of absorption
lines in some LRDs suggest a BAL origin in at least some objects.
Metastable helium (\ion{He}{2}$*\lambda 10830$) absorption has been found
in at least one LRD \citep{wang25}. First reported in a BAL quasar by
\citet{leighly11}, it is now known to be common among LoBAL quasars
\citep{liu15}, and is predicted to be present in many of the
low-redshift FeLoBAL quasars that we studied \citep{choi22}.  Second,
it has been noticed that in LRDs the ratio of the opacity in H$\alpha$
to H$\beta$ is smaller than required by atomic physics, just like it is
in the FeLoBAL quasars presented here.  Though this may be evidence
for a stellar atmosphere explanation 
\citep[e.g.][]{deugenio25}, it is possible that partial covering is
common in LRDs, just like it is in BAL quasars.  

If the origin of the Balmer absorption lines in LRDs is similar to the
origin in FeLoBAL quasars, then where is the accompanying FeLoBAL
absorption?  It may be present but hidden.  First, LRDs are defined by
a steeply reddened rest-frame optical spectrum.  FeLoBAL absorption
occurs shortward of \ion{Mg}{2}$\lambda 2800$ chiefly, so that part of
the spectrum may simply be too faint to see \ion{Fe}{2} absorption.
Second, LRDs are also found to 
have a blue spectrum at short wavelengths, which is likely to be a
separate component.  That separate component may fill in the absorbed
and reddened spectrum, making the \ion{Fe}{2} absorption lines too
shallow to detected \citep[e.g.,][]{li21}.  

\subsubsection{Emission Lines} \label{emission}

High-redshift quasars typically have Lorenzian-shaped Balmer lines and
strong optical \ion{Fe}{2} emission \citep[e.g.,][]{temple24}.  While
LRDs are identified by their broad Balmer emission lines, their
emission lines are much different.  They tend to have weak or absent
\ion{Fe}{2} \citep{trefoloni24}.  Their Balmer lines tend to have both
a narrow and broad component, i.e., like
intermediate-type\footnote{An empirical classification,
intermediate-type Seyfert galaxies fall between Seyfert 1 galaxies
which have no detectable narrow line region contribution to their
Balmer lines, and Seyfert 2 galaxies in which only the narrow line region
emission is observed \citep{osterbrock81}.}  Seyfert 
galaxies \citep[e.g.,][]{stern12b}.  The LRDs are found to have
broader H$\alpha$ lines, and a higher broad-to-narrow H$\alpha$ ratio
than other high redshift H$\alpha$ selected AGN \citep{taylor25}.

At low redshift, the \ion{Fe}{2}, H$\beta$, and [\ion{O}{3}]
properties are historically related by a construct known as
Eigenvector 1; namely, there is known to be a set of correlations between
\ion{Fe}{2} and [\ion{O}{3}] strengths and H$\beta$ widths that is
repeatedly found in rest-frame quasar spectra \citep[e.g.,][]{bg92,
  sulentic00, grupe04, ludwig09, wolf20}.  This pattern of emission
line behavior is further linked with the quasar Eddington ratio
\citep[e.g.,][]{shen_ho_14}.  It was originally thought that BAL
quasars were  high accretion rate objects, i.e., generally
showing strong \ion{Fe}{2} and weak [\ion{O}{3}], and relatively
narrow Balmer lines \citep{boroson02}.  However, that scenario was
shown to be wrong for low-redshift FeLoBAL quasars by
\citet{leighly22}.  Instead, we found that FeLoBAL quasars were
characterized by high and low Eddington ratios, with a lack of objects
with intermediate Eddington ratios.

The low-redshift loitering outflow FeLoBAL quasars in
\citet{leighly22} are all members of the low accretion rate class.
Many of the objects show intermediate-type rest-frame optical spectra;
one object was originally classified as a Type 2 quasar \citep{yuan16}.
The composite spectrum from the low accretion rate FeLoBAL quasars 
\citep[][Fig.\ 9]{leighly22} shares the weak \ion{Fe}{2}, strong
[\ion{O}{3}] emission, and intermediate classification
characteristic of many LRDs.

On the other hand, the spectra presented in this paper and shown in
Fig.~\ref{specfit_plots} are mostly different than those from LRDs.
These differences might be attributed to the different ranges of
luminosity represented, as many quasar properties are known to be
luminosity and accretion rate dependent.  First, like X-ray binaries,
the configuration of the quasar central engine is believed to change 
dramatically as the accretion rate is dialed up \citep[e.g.,][]{gp19,
  hagen24, hopkins25}.  These changes are thought to produce a
different broad-band spectral energy distributions, and they may also be
accompanied by changes in the broad-line region emission and
configuration \citep[e.g.,][Fig.\ 15]{leighly04}.  Second, in
flux-limited samples like the SDSS, higher-redshift objects inevitably
have higher luminosities, and because the dynamic range of luminosity is
larger than that of Eddington ratio, it means that at higher redshift, 
low accretion rate objects drop out of the sample \citep{jester05}.  

JWST affords the first thorough look at lower-luminosity objects at
high redshifts.  LRDs have estimated bolometric luminosities between
$45 < \log_\mathrm{10} L_\mathrm{Bol} < 47$ [erg s$^{-1}$]
\citep{akins24}.  The bolometric luminosities of the loitering outflow
FeLoBAL quasars in this paper range between
$46.3 < \log_\mathrm{10} L_\mathrm{Bol} < 47.5$  [erg s$^{-1}$], i.e.,
overlapping but higher than many LRDs.  However, the low accretion 
rate FeLoBALs in the low redshift sample presented in
\citet{leighly22} range between $\sim 45.5$ and $\sim 46.4$, i.e.,
more typical of the bolometric luminosity estimates for LRDs.  We
therefore suggest that, to some extent, LRDs have different optical
spectra than high-redshift quasars because their luminosity is lower
than typical high-redshift quasars. 

\subsubsection{Spectral Shape} \label{shape}

One of the characteristic properties of LRDs is that they have a
V-shaped spectrum \citep[e.g.,][]{kocevski25}, heavily reddened in
their rest-frame optical band, with a recovery to a blue spectrum in
the near-UV.  It is not clear that any of the standard extinction
curves satisfactorily model the reddening \citep[e.g.,][]{ma25,
  chen25}.  For example, \citet{li25, ma25} suggested that a
deficit of small dust grains, producing a grey extinction curve in the UV, might
explain the spectra.  

Generally, BAL quasars show stronger reddening and a
higher scattering fraction than unabsorbed quasars
\citep[e.g.,][]{sf92, brotherton97,   dipompeo11, krawczyk15}.  Thus
their spectra are in some cases also somewhat V-shaped, although they
are different in their details from LRDs: BALQ optical spectra can be
quite blue at long wavelengths with the steep reddening at shorter
wavelengths starting 3000-5000\AA, and the recovery to a flat
spectrum, when it is present, occurring shortward of $\sim 1800-2500$\AA\/
\citep[e.g.,][]{leighly09,leighly14}.  The steep reddening curve,
known as ``anomalous reddening'', does not resemble any of the
commonly used extinction curves from the Milky Way or Magellanic
clouds \citep[e.g.][]{ccm88, prevot84}; rather, the curves are very
steep through near-UV \citep{hall02, leighly09, fynbo13,   jiang13,
  leighly14,  krogager15, meusinger16}.  Anomalous reddening is rare,
but it appears to be more common in FeLoBALQs \citep[six objects among
  $\sim 50$][require it]{choi20, choi22}.  In the case of
FBQS~1408$+$3054, the reddening can be modeled well using a lognormal 
dust distribution with mean and standard deviation of $246$ and $0.15$
microns respectively (Choi et al.\ in prep.).   

The origin of the blue spectrum shortward of the near UV in LRDs is
not known, but at least in some cases it is consistent with scattering
\citep[e.g.,][]{greene24}.  Evidence for scattered light is common
in BALQ spectra.  BALQs are often strongly polarized in their rest-UV
spectra \citep[e.g.,][]{ogle99, dipompeo10, dipompeo11, dipompeo13};
the polarization properties suggest electron or dust scattering 
\citep[e.g.,][]{wills92, leighly97}.  In addition, it is well known
that partial covering is required to explain the observed optical
depths BAL quasar spectra \citep{hamann98a}.  The origin of partial
covering is not known \citep[e.g.,][]{leighly19}; in some cases it is
likely to arise from the details of the opacity
\citep[e.g.,][]{green23}, but in others, scattering may be important
\citep[e.g.,][]{choi20,choi22}.

\subsubsection{Hot Dust Emission}\label{hotdust}

One of the puzzling properties of LRDs the lack of hot dust emission;
specifically, the spectral energy distributions do not show the
typical flattening longward of one micron due to hot
($T_\mathrm{dust}\sim 1500\, \rm K$)  dust emission from the inner
edge of the torus \citep[e.g.,][]{akins24, wang25, chen25}.

Luminous high-ionization broad absorption line quasars have spectral
energy distributions, including torus emission, indistiguishable from 
unbsorbed quasars \citep[e.g.,][]{gallagher07b, ahmed25}.  In fact,
evidence for stronger torus emission in BALQs compared with unabsorbed 
quasars has been found in some objects \citep{dipompeo13b}.

\citet{leighly24} studied the hot dust properties in a sample of
low-redshift FeLoBAL quasars along with a redshift- and
luminosity-matched unabsorbed comparison sample.  We found a strong
correlation  between the Eddington ratio and the strength of the torus
emission among the unabsorbed quasars.  We also found that the
low-redshift loitering outflow objects are significantly weak in hot
dust emission. 
\citet{leighly24} linked these 
correlations to the predicted behavior of the torus as a function of
accretion rate in torus wind models.  Specifically, it has been
proposed that large-scale dynamical outflows are present in quasars,
and the torus may be the portion of the wind that is optically thick
enough to both block 
the view of the central engine in AGN unified models and to
thermalize incident radiation into emission in the infrared
\citep[e.g.,][]{bp82, kk94, gallagher15}.  Building upon this idea,
\citet{es06} suggested that the torus should disappear at low enough
accretion rates that the optically thick wind cannot be
sustained. \citet{eh09} take these ideas one step further, suggesting
that at sufficiently low accretion rates, the broad-line region should 
disappear.  \citet{leighly24} notes that there is substantial evidence
in the literature that low-accretion-rate active galaxies lack a torus
and may also lack a broad-line region (see \S~8.1 in that paper).   

\subsubsection{X-ray Properties}\label{xray}

LRDs have been observed to be X-ray weak \citep[e.g.,][]{akins24,
  wang25, yue24,   kocevski25}.  A few objects have been X-ray
detected, and their X-ray spectra reveal moderate column densities
\citep[e.g.,][]{kocevski25}.

BAL quasars are known to be generally X-ray weak
\citep[e.g.,][]{green01}.  Among high-ionization broad absorption
line quasars, X-ray spectral analysis generally shows evidence for
X-ray absorption; correcting the spectra for this absorption often
recovers a normal UV-to-X-ray flux ratio \citep{gallagher02b,
  gallagher06}.  However, some special classes BAL quasars have
minimal X-ray absorption \citep[e.g.,][]{brotherton05, giustini08,
  miller09, gibson09b}, while 
LoBAL quasars and FeLoBAL quasars have more X-ray absorption
\citep{green01, gallagher06, morabito11}.  Hard X-ray observations of
some BAL quasars show them to be intrinsically X-ray weak
\citep{luo13,luo14}.  Intrinsic X-ray weakness was also found in the
bright, nearby quasar PHL~1811; the X-ray weakness was attributed to
the lack of an X-ray emitting corona \citep{leighly07a}.  Such
intrinsic X-ray weakness has been found among JWST AGN, and a similar
explanation posited \citep{maiolino25}.  

\section{Summary and Conclusions}  \label{summary}

We presented analysis of a sample of 14 FeLoBAL quasars that exhibit
Balmer absorption in their rest-frame optical spectra.  These include
a 9-object subsample from our near-infrared spectroscopic observations 
of high-redshift (0.97--2.58) FeLoBALQs, supplemented by one archival
observation, two spectra from the literature, and two spectra from the
SDSS that don't cover H$\alpha$ but do show H$\beta$ absorption.
Eight of these spectra reveal new detections of Balmer absorption 
lines.  Using spectral fitting (Fig.~\ref{specfit_plots}), we extracted
the offset velocity, velocity width, and optical depth of the Balmer
absorption lines.  We computed the apparent H(n=2) column density.
Taking account of partial covering and the wide range of
signal-to-noise ratios of our spectra, we also estimated the true
column density (Table~\ref{specfit_tab}).  We estimated the quasar
accretion properties including the black hole mass, bolometric
luminosity, and Eddington ratio.  Finally, we estimated $\alpha_{oi}$,
the  point-to-point slope between 5100\AA\/ and 3$\mu\rm m$
(Table~\ref{global_tab}). 

We performed a correlation analysis among the measured parameters
(Fig.~\ref{correlation_plots}).  We found an
anticorrelation between Eddington ratio and outflow velocity
(Fig.~\ref{velcor_plot}).  Generally found among BAL quasars, the
correlation is weak in this sample because nearly all objects are
loitering outflow FeLoBAL quasars \citep{choi22}, identified by their
low outflow velocity and compact absorption.  We found
anticorrelations between $L_\mathrm{Bol}$ or Eddington ratio and
various computed parameters measuring $dN/dv$, i.e., the column
density per unit velocity (Fig.~\ref{dndvcorr_plot}).  Finally, we
found a correlation between $\alpha_{oi}$ and the absorption velocity  
width (Fig.~\ref{fwhm_vs_alphaoi_plot}).

We discussed the implications of these results on our understanding of
FeLoBAL quasars (\S~\ref{properties}).  Among low-redshift FeLoBAL
quasars, \citet{choi22} predicted that nearly all objects with compact
($\log R < 1 \rm \, [pc]$) outflows should have Balmer absorption
lines, including both loitering outflow FeLoBALQs and overlapping
trough quasars, distinguishable by their dramatically different
kinematics.  In this paper, we presented only the objects that show 
Balmer absorption; this criterion includes all the loitering outflow
objects, but excludes all the overlapping trough objects, likely
because their absorption troughs are so broad that the absorption
blends into the continuum. Thus, there is an important selection
effect influencing the detectability of Balmer absorption.

Assuming that the outflow location and total column density are the
same for all of the detected objects, we computed measures proportional
to the mass outflow rate $\dot M_\mathrm{out}$, momentum flux $\dot
p$, and kinetic luminosity $L_\mathrm{KE}$.  In contrast with other
outflow samples, we found that the mass outflow rate does not depend
on $L_\mathrm{Bol}$ or Eddington ratio, while both $\dot p$ and
kinetic luminosity $L_\mathrm{KE}$ are positively correlated.  We
noted, however, that the spread in $\dot M_\mathrm{out}$ is very large,
and we will revisit this result using the {\tt SimBAL} analysis
results (Choi et al.\ in prep.).  

Finally, we compared the properties of FeLoBAL quasars with the
Little Red Dots, a type of object discovered at high redshift by JWST
(\S~\ref{LRD}).  We do not claim that FeLoBAL quasars are a local
analog of LRDs, but there are several suggestive similarities that may
point to common physical origins of like phenomena.  The first is the
presence of Balmer absorption (\S~\ref{balmer}).  LRDs and FeLoBAL
quasars  are essentially the only types of active galaxies that show
Balmer absorption, and because the physical conditions for Balmer
absorption are so extreme, it is possible that they may be the same in
some cases.  LRDs do not show FeLoBAL absorption lines, but those
could be hidden by the strong reddening and blue short wavelength
continuum that is their hallmark.

LRDs are also similar in their emission line properties to the low
accretion rate branch of FeLoBALs described by \citet{leighly22} in
that they often have intermediate Seyfert class Balmer lines,  lack
\ion{Fe}{2} emission, and have strong [\ion{O}{3}] lines, i.e., much
different than the typical high-redshift quasar
(\S~\ref{emission}). We suggested  that this is a consequence of their
lower luminosity compared with high-redshift quasars.  We noted the
similarities in spectral shape, 
specifically the steep reddening that might be related to the
anomalous reddening observed in some FeLoBALQs, and blue continuum
(\S~\ref{shape}) that, being polarized, has a scattering origin in
BALQs. We noted the lack of hot dust in LRDs and loitering outflow
FeLoBAL quasars (\S~\ref{hotdust}).  For FeLoBAL quasars, we have
suggested that this is a consequence of their low accretion rate
\citep{leighly24}.   Finally, we noted that both classes are weak in
their X-ray emission (\S~\ref{xray}).  

Observations of a common property in a set of objects does not mean
that all objects in that set are in all ways identical.  For
example, all FeLoBAL quasars have \ion{Fe}{2} absorption, but
\citet{leighly22} showed that they are divided into high and low
accretion rate objects.  Likewise, the similarities between
loitering outflow FeLoBAL quasars and LRDs does not mean that they are 
the same.  Nevertheless, the patterns  discussed in this
paper, spanning many observational properties, may offer some clues.

\begin{acknowledgments}
KML acknowledges useful discussion with Chanuntorn Pumpo, Gilberto
Garcia, and Jorge Escalera.  Support for {\tt SimBAL} development and
analysis is provided by NSF Astronomy and Astrophysics Grants
No.\ 1518382, 2006771, and 2007023.  S.C.G. acknowledges the support
of the Natural Science and Engineering Research Council
(RGPIN-2021-04157) and a Western University Research Leadership Chair
Award. LKM is grateful for support from UKRI [MR/Y020405/1] 

Long before the University of Oklahoma was established, the land on
which the University now resides was the traditional home of the
“Hasinais” Caddo Nation and “Kirikiris” Wichita \& Affiliated
Tribes. This land was also once part of the Muscogee Creek and
Seminole nations.

We acknowledge this territory once also served as a hunting ground,
trade exchange point, and migration route for the Apache, Comanche,
Kiowa and Osage nations. Today, 39 federally-recognized Tribal nations
dwell in what is now the State of Oklahoma as a result of settler
colonial policies designed to assimilate Indigenous peoples.

The University of Oklahoma recognizes the historical connection our
university has with its Indigenous community. We acknowledge, honor
and respect the diverse Indigenous peoples connected to this land. We
fully recognize, support and advocate for the sovereign rights of all
of Oklahoma’s 39 tribal nations.

This acknowledgement is aligned with our university’s core value of
creating a diverse and inclusive community. It is our institutional
responsibility to recognize and acknowledge the people, culture and
history that make up our entire OU Community.

Funding for the SDSS and SDSS-II has been provided by the Alfred
P. Sloan Foundation, the Participating Institutions, the National
Science Foundation, the U.S. Department of Energy, the National
Aeronautics and Space Administration, the Japanese Monbukagakusho, the
Max Planck Society, and the Higher Education Funding Council for
England. The SDSS Web Site is http://www.sdss.org/.

The SDSS is managed by the Astrophysical Research Consortium for the
Participating Institutions. The Participating Institutions are the
American Museum of Natural History, Astrophysical Institute Potsdam,
University of Basel, University of Cambridge, Case Western Reserve
University, University of Chicago, Drexel University, Fermilab, the
Institute for Advanced Study, the Japan Participation Group, Johns
Hopkins University, the Joint Institute for Nuclear Astrophysics, the
Kavli Institute for Particle Astrophysics and Cosmology, the Korean
Scientist Group, the Chinese Academy of Sciences (LAMOST), Los Alamos
National Laboratory, the Max-Planck-Institute for Astronomy (MPIA),
the Max-Planck-Institute for Astrophysics (MPA), New Mexico State
University, Ohio State University, University of Pittsburgh,
University of Portsmouth, Princeton University, the United States
Naval Observatory, and the University of Washington.

Funding for SDSS-III has been provided by the Alfred P. Sloan
Foundation, the Participating Institutions, the National Science
Foundation, and the U.S. Department of Energy Office of Science. The
SDSS-III web site is http://www.sdss3.org/.

SDSS-III is managed by the Astrophysical Research Consortium for the
Participating Institutions of the SDSS-III Collaboration including the
University of Arizona, the Brazilian Participation Group, Brookhaven
National Laboratory, Carnegie Mellon University, University of
Florida, the French Participation Group, the German Participation
Group, Harvard University, the Instituto de Astrofisica de Canarias,
the Michigan State/Notre Dame/JINA Participation Group, Johns Hopkins
University, Lawrence Berkeley National Laboratory, Max Planck
Institute for Astrophysics, Max Planck Institute for Extraterrestrial
Physics, New Mexico State University, New York University, Ohio State
University, Pennsylvania State University, University of Portsmouth,
Princeton University, the Spanish Participation Group, University of
Tokyo, University of Utah, Vanderbilt University, University of
Virginia, University of Washington, and Yale University.

\end{acknowledgments}

\begin{contribution}

\textbf{Karen Leighly:} conceptualization; data curation; formal
analysis; funding acquisition; investigation; methodology; project
administration; resources; software; visualization; writing --
original draft, \textbf{Sarah Gallagher:} conceptualization;
investigation; writing - review and editing, \textbf{Donald Terndrup:} 
conceptualization; funding acquisition; investigation; software; 
writing - review and editing, \textbf{Hyunseop Choi:}
conceptualization; funding acquisition; methodology;  resources;
software; writing - review and editing. \textbf{Julianna R.\ Voelker:}
writing - review and editing. \textbf{Gordon Richards:} 
writing - review and editing.  \textbf{Leah K.\ Morabito:} writing -
review and editing. 


\end{contribution}

%
\facilities{Gemini(GNIRS), APO (Triplespec)}

\software{Sherpa \citep{sherpa24},  
          Cloudy \citep{ferland17}, SimBAL \citep{leighly18},
          PypeIt \citep{prochaska20}, Triplespectool \citep{cushing04,
            vacca03} 
          }





\bibliographystyle{aasjournalv7}



\end{document}